\documentstyle[fleqn,12pt]{article}
\renewcommand{\theequation}{\arabic{section}.\arabic{equation}}

\newcommand{\ra}{\rightarrow}
\newcommand{\bra}{\langle}
\newcommand{\ket}{\rangle}
\newcommand{\be}{\begin{equation}}
\newcommand{\ee}{\end{equation}}
\newcommand{\bea}{\begin{eqnarray}}
\newcommand{\eea}{\end{eqnarray}}
\newcommand{\eps}{\varepsilon}
\newcommand{\ep}{\hfill {$\Box$}}
\newcommand{\e}{\mbox{e}}

\newcommand{\ffi}{\varphi}
\newcommand{\ode}{{\cal O}}

\newtheorem{lem}{Lemma}[section]
\newtheorem{prop}{Proposition}[section]
\newtheorem{thm}{Theorem}[section]

\newcommand{\R}{{\bf R}}
\newcommand{\N}{{\bf N}}
\newcommand{\C}{{\bf C}}

\newcommand{\un}{{\bf I}}
\newcommand{\zer}{{\bf O}}

\begin{document}

\title{Exponential Asymptotics in a Singular Limit for $n$-Level
Scattering Systems}

\author{Alain Joye \\
Centre de Physique Th\'eorique \\
C.N.R.S. Marseille\\
Luminy Case 907\\F-13288 Marseille Cedex 9\\
and \\
Phymat, Universit\'e de
Toulon et du Var\\
B.P. 132\\ 83957 La Garde Cedex  \date{June 28, 1995} }

\maketitle
\abstract{The singular limit $\eps\ra 0$ of the $S$-matrix
associated with the equation $i\eps d\psi(t)/dt=H(t)\psi(t)$ is considered,
where the analytic generator $H(t)\in M_n(\C)$ is such that its spectrum
is real and non-degenerate for all $t\in\R$. Sufficient conditions
allowing to compute asymptotic formulas for the exponentially small
off-diagonal elements of the $S$-matrix as $\eps\ra 0$ are explicited and
a wide class of generators for which these conditions are verified is defined.
These generators are obtained by means of generators whose spectrum
exhibits
eigenvalue crossings which are perturbed in such a way that these
crossings
turn to avoided crossings. The exponentially small asymptotic formulas
which
are derived are shown to be valid up to exponentially small relative
error,
by means of a joint application of the complex WKB method together with
superasymptotic renormalization. The application of these results to the
study
of quantum adiabatic transitions in the time dependent Schr\"odinger
equation
and of the semiclassical scattering properties of the multichannel
stationary Schr\"odinger equation closes this paper. The results presented
here are a generalization to $n$-level systems, $n\geq 2$, of results
previously known for $2$-level systems only.
}
\newline
\vspace{2cm}\newline
CPT-95/P.3216
\\ e-mail: joye@cpt.univ-mrs.fr

\newpage
\section{Introduction}
\setcounter{equation}{0}

Several problems of mathematical physics lead to the study of the
scattering
properties of linear ordinary differential equations in a singular limit
\be\label{sch}
  i\eps \psi'(t)=H(t)\psi(t),\;\;\; t\in\R, \;\;\eps\ra 0,
\ee
where the prime denotes the derivative with respect to $t$,
$\psi(t)\in\C^n$,
$H(t)\in M_n(\C)$, for all $t$. A system described by such an equation
will be called a
$n$-level system. Let us mention for example the study of the adiabatic
limit of the time dependent Schr\"odinger equation or the semiclassical
limit of the
one-dimensional multichannel stationary Schr\"odinger equation at
energies above
the potential barriers, to which we will come back below. When the
generator $H(t)$
is well behaved at $+\infty$ and $-\infty$, the scattering properties of
the problem
can be described by means of a matrix naturally associated with equation
(\ref{sch}),
the so-called $S$-matrix. This matrix relates the behavior of
the solution $\psi(t)$
as $t\ra -\infty$ to that of $\psi(t)$ as $t\ra +\infty$. Assuming that
the spectrum
$\sigma(t)$ of $H(t)$ is real and non-degenerate,
\be
  \sigma(t)=\{e_1(t)<e_2(t)<\cdots <e_n(t)\}\in \R,
\ee
the $S$-matrix is essentially given by the identity matrix
\be
  S=\mbox{diag }(s_{11}(\eps),s_{22}(\eps),\cdots s_{nn}(\eps)) +
  \ode(\eps^{\infty}),\;\;\;
  \mbox{where }s_{jj}(\eps)=1+\ode(\eps)\mbox{ as }\eps\ra 0,
\ee
provided $H(t)$ is $C^{\infty}$, see e.g. \cite{f1}, \cite{f2}, \cite{w}.
Moreover, if $H(t)$ is assumed to
be analytic, it was proven in various situations that the off-diagonal
elements
$s_{jk}$ of $S$ are exponentially decreasing \cite{ff}, \cite{w},
\cite{f1}, \cite{f2}, \cite{jkp}, \cite{jp3}
\be
  s_{jk}=\ode \left(\e^{-\kappa /\eps}\right),\;\;\; \forall j\neq k,
\ee
as $\eps\ra 0$.
See also \cite{jp4}, \cite{n}, \cite{m}, \cite{sj} for corresponding
results in infinite dimensional spaces.
Since the physical information is often contained in these off-diagonal
elements, it is of interest to be able to give an asymptotic formula for
$s_{jk}$, rather
then a mere estimate.

For $2$-level systems (or systems reducible to this case, see \cite{jp5},
\cite{j1}, \cite{mn}), the situation is now
reasonably well understood, at least under generic circumstances. Indeed,
a rigorous study of the $S$-matrix associated with (\ref{sch}) when $n=2$,
under the hypotheses
loosely stated above, is provided in the recent paper \cite{jp3}. The
treatment presented unifies
in particular earlier results obtained either for the time dependent
adiabatic schr\"odinger equation, see e.g \cite{jp7} and references
therein, or for
the study of the above barrier reflexion in the semiclassical limit, see
e.g. \cite{ff},\cite{o}.
Further references are provided in \cite{jp3}. As an intermediate result,
the asymptotic formula
\be\label{exaf}
  s_{jk}=g_{jk}\e^{-\Gamma_{jk}/\eps}\left(1+\ode (\eps)\right),\;\;\;\eps
  \ra 0,
\ee
for $j\neq k\in\{1,2\}$, with $g_{jk}\in\C$ and $\mbox{Re }\Gamma_{jk}>0$
is proven in \cite{jp3}.
As is well known, to get an asymptotic formula for $s_{jk}$, one has to
consider (\ref{sch}) in the complex plane, in particular in the vicinity
of the degeneracy points of the
analytic continuations of eigenvalues
$\e_1(z)$ and $e_2(z)$. {\em Provided} the level lines of the multivalued
function
\be\label{lln}
  \mbox{Im }\int_0^ze_1(z')-e_2(z')dz'= \mbox{cst},
\ee
called Stokes lines, naturally associated with (\ref{sch}) behave
properly in the
complex plane, the so-called complex WKB method allows to prove
(\ref{exaf}).
But more important, it is also shown in \cite{jp3} how to improve
(\ref{exaf})
to an asymptotic formula accurate up to exponentially small relative
 error
\be\label{exasup}
  s_{jk}=g_{jk}^*(\eps)\e^{-\Gamma_{jk}^*(\eps)/\eps}\left(1+\ode
  \left(\e^{-\kappa /\eps}\right)\right),
  \;\;\;\eps\ra 0,
\ee
with $g_{jk}^*(\eps)=g_{jk}+\ode (\eps)$ and $\Gamma_{jk}^*(\eps)=
\Gamma_{jk}+\ode(\eps^2)$.
This is achieved by
using a complex WKB analysis jointly with the recently developed
superasymptotic theory
\cite{be}, \cite{n}, \cite{jp5}.
Note that when given a generator,
the principal difficulty in justifying formulas (\ref{exaf}) and
(\ref{exasup}) is the verification
that the corresponding Stokes
lines (\ref{lln}) display the proper behavior {\em globally} in the
complex plane, which may or may not be the case
\cite{jkp}. However, this condition is always satisfied when the complex
eigenvalue degeneracy is close
to the real axis, as shown in \cite{j1}. See also \cite{mn} and \cite{r}
for recent related results.

For $n$-level systems, with $n\geq 3$, the situation is by no means as
well understood.
There are some results obtained with particular generators. In \cite{de},
\cite{ch1},
\cite{ch2} and \cite{bve}, certain elements of the $S$-matrix are
computed if $H(t)=H^*(t)$ depends
linearly on $t$, $H(t)=A+tB$,
for some particular matrices $A$ and $B$. The choices of $A$ and $B$ are
such that all
components of the solution
$\psi(t)$ can be deduced from the first one and an exact integral
representation of this
first component can be obtained. The integral representation is analyzed
by standard asymptotic
techniques and this leads to results which are valid for any $\eps>0$, as
for the classical
Landau-Zener generator.
The study of the three-level problem when $H(t)=H^*(t)\in M_3(\R)$ is
tackled in the closing
section of the very interesting paper \cite{hp}.
A non rigorous and essentially local discussion
of the behavior of the level lines $\mbox{Im }\int_0^ze_j(z')-e_k(z')dz'$,
$j\neq k=1,2,3$, is provided
and justifies in very favorable cases an asymptotic formula for some
elements of the
$S$-matrix. See also the review \cite{s}, where a non rigorous study of
(\ref{sch})
is made close to a complex degeneracy point of a group of eigenvalues by
means of an exact solution
to a model equation.
However, no asymptotic formula for $s_{jk}$, $j\neq k$, can be
found in the literature for general $n$-level systems, $n\geq 3$. This is
due to the fact
that the direct generalization of the method used successfully for
$2$-level
systems may lead to seemingly inextricable difficulties for $n=3$
already. Indeed, with three eigenvalues,
one has to consider three sets of level lines $\mbox{Im }
\int_0^ze_j(z')-e_k(z')dz'$ to
deal with (\ref{sch}) in the complex plane, and the conditions they have
to fulfill in order that
the limit $\eps\ra 0$ can be controlled may
be incompatible for a given generator, see \cite{f1}, \cite{f2} and
\cite{hp}. It should be mentioned however, that
there are specific examples in which this difficult problem can be
mastered. Such a result
was recently obtained in the semiclassical study \cite{ba} of a particular
 problem of
resonances for which similar considerations in the complex plane are
required.

The goal of this paper is to provide some general insight on the
asymptotic computation of the
$S$-matrix associated with $n$-level systems, $n\geq 3$, based on a
generalization of the techniques which proved
to be successful for $2$-level systems.
The content of this paper is twofold. On the one hand we set up a general
framework in which
asymptotic formulas for the exponentially small off-diagonal coefficients
can be proven. On the other hand we actually prove such formulas for a
wide class of $n$-level systems.
In the first part of the paper, we give our definition of the $S$-matrix
associated with
equation (\ref{sch}) and explicit the symmetries it inherits from the
symmetries of $H(t)$,
for $t\in\R$ (proposition \ref{PROS}). Then we
turn to the determination of the analyticity properties of the
eigenvalues
and eigenvectors of $H(z)$, $z\in\C$, which are at the root of the
asymptotic formulas we derive
later (lemma \ref{ECHCO}). The next step is the formulation of sufficient
conditions adapted to the scattering situation
we consider, under which a complex WKB analysis allows to prove a formula
like (\ref{exaf})
(proposition \ref{eslf}). The
conditions stated are similar but not identical to those given in
\cite{jkp} or \cite{hp}.
As a final step, we show how to improve the asymptotic formula
(\ref{exaf}) to (\ref{exasup}) by means of the superasymptotic machinery
(proposition \ref{MAPROP} and
lemma \ref{sdec}).
Then we turn to the second part of the paper, where we show that a wide
class of
generators fits into our framework and satisfies our conditions.
These generators are obtained by perturbation of
generators whose eigenvalues display degeneracies on the real axis, in
the spirit of \cite{j1}.
We prove that for these generators, in absence of any symmetry of the
generator $H(t)$, one element
per column at least in the $S$-matrix can be
asymptotically computed (theorem \ref{PERCO}).
This is the main technical section of the paper. The major advantage of
this
construction is that it is sufficient to look at the behavior of the
eigenvalues on the real axis to
check if the conditions are satisfied. The closing section contains an
application of our general
results to the study of quantum adiabatic transitions in the time
dependent Schr\"odinger equation
and of the semiclassical scattering properties of the multichannel
stationary Schr\"odinger equation. In particular, explicit use of the
symmetries of the $S$-matrix is
made to increase the number of its elements for which an asymptotic
formula holds.
In the latter application, further specific symmetry properties
of the $S$-matrix are derived (lemma \ref{REPOT}).
\vspace{.5cm} \\
{\bf Acknowledgments:} It is a great pleasure to thank Charles-Edouard
Pfister
for many enlightening and fruitful discussions which took place in
Marseille and
Lausanne. The hospitality of the Institut de Physique Th\'eorique de
l'EPFL
where part of this work was accomplished is acknowledged.

\section{Definition and properties of the $S$-matrix.}
\setcounter{equation}{0}

We consider the evolution equation
\be\label{schr}
  i\eps \psi'(t)=H(t)\psi(t),\;\;\; t\in\R, \;\;\eps\ra 0,
\ee
where the prime denotes the derivative with respect to $t$,
$\psi(t)\in\C^n$,
$H(t)\in M_n(\C)$, for all $t$.
We make some assumptions on the
generator $H(t)$. The first one is the usual analyticity condition in this
context.\\
{\bf H1}{\em
There exists a strip
\be
  S_{\alpha}=\left\{z\in\C |\, |\mbox{\em Im }z|\leq\alpha\right\},\;\;
  \alpha>0,
\ee
such that $H(z)$ is analytic for all $z\in S_{\alpha}$.}\\ Since we are
studying scattering properties, we need sufficient decay at infinity.\\
{\bf H2} {\em There exist two nondegenerate matrices $H(+), H(-) \in
 M_n(\C)$
and $a>0$ such that
\be
  \lim_{t\ra\pm \infty}|t|^{1+a}\sup_{|s|\leq\alpha}\|H(t+is)-H(\pm)\|
  <\infty.
\ee}
We finally give a condition which has to do with the physics behind
the problem.\\
{\bf H3} {\em
For $t\in\R$, the spectrum of $H(t)$, denoted by $\sigma(t)$,
is real and non-degenerate
\be
  \sigma(t)=\left\{e_1(t)<e_2(t)<\cdots <e_n(t)\right\}\subset \R
\ee
and there exists $g>0$ such that
\be
  \inf_{j\neq k \atop t\in \R}\left|e_j(t)-e_k(t)\right|\geq g.
\ee
}
As a consequence of H3, for each $t\in \R$, there exists a complete set
of projectors $P_j(t)=P_j^2(t)\in M_n(\C)$, $j=1,2,\cdots ,n$ such that
\bea
  & &\sum_{j=1}^n P_j(t)\equiv \un\\
  & &H(t)=\sum_{j=1}^n e_j(t)P_j(t)
\eea
and there exists a basis of $\C^n$ of eigenvectors of
$H(t)$. We determine these eigenvectors $\ffi_j(t)$, $j=1,2,\cdots ,n$
uniquely (up to a constant) by requiring them to satisfy
\bea
  & &H(t)\ffi_j(t)=e_j(t)\ffi_j(t)\\
  & &P_j(t)\ffi_j'(t)\equiv 0, \;\;\;\;\;\;\;\;\;\;j=1,2,\cdots ,n.
  \label{phco}
\eea
Explicitly, if $\chi_j(t)$, $j=1,2,\cdots ,n$ form a complete set of
differentiable eigenvectors of $H(t)$, the eigenvectors
\be
 \ffi_j(t)=\e^{-\int_0^t\xi_j(t')dt'}\psi_j(t),\;\;\;\mbox{s.t. }\ffi_j(0)
 =\psi_j(0)
\ee
with
\be
  \xi_j(t)=\frac{\bra\psi_j(t)|P_j(t)\psi_j'(t)\ket}{\|\psi_j(t)\|^2},
  \;\;\;
  j=1,\cdots ,n
\ee
verify (\ref{phco}).
That this choice leads to an analytic set of eigenvectors close to the
 real axis will be proven below.
We expand the solution $\psi(t)$ along the basis just constructed, thus
defining unknown coefficients $c_j(t)$, $j=1,2,\cdots ,n$ to be determined,
\be\label{expa}
  \psi(t)=\sum_{j=1}^nc_j(t)\e^{-i\int_0^te_j(t')dt'/\eps }\ffi_j(t).
\ee
The phases $\e^{-i\int_0^te_j(t')dt'/\eps }$ (see H3) are introduced for
convenience. By inserting (\ref{expa}) in (\ref{schr}) we get the
following differential equation for the $c_j(t)$'s
\be\label{eqco}
  c_j'(t)=\sum_{k=1}^na_{jk}(t)\e^{i\Delta_{jk}(t)/\eps}c_k(t)
\ee
where
\be
  \Delta_{jk}(t)=\int_0^t(e_j(t')-e_k(t'))dt'
\ee
and
\be\label{coup}
  a_{jk}(t)=-\frac{\bra\ffi_j(t)|P_j(t)\ffi_k'(t)\ket}{\|\ffi_j(t)\|^2}.
\ee
Here $\bra \cdot |\cdot \ket$ denotes the usual scalar product in $\C^n$.
Our choice (\ref{phco}) implies $a_{jj}(t)\equiv 0$. It is also
shown below that the $a_{jk}(t)$'s are analytic functions in a
neighborhood
of the real axis and that hypothesis H2 imply that they satisfy the
estimate
\be
  \lim_{t\ra\pm\infty}\sup_{j\neq k}
  |t|^{1+a}\left|a_{jk}(t)\right|<\infty.
\ee
As a consequence of this last property and of the fact that the
eigenvalues
are real by assumption, the following limits exist
\be
  \lim_{t\pm\infty}c_j(t)=c_j(\pm\infty).
\ee
We are now able to define the associated $S$-matrix, $S\in M_n(\C)$, by
the
identity
\be
  S\pmatrix{c_1(-\infty)\cr c_2(-\infty) \cr \vdots \cr c_n(-\infty)}=
  \pmatrix{c_1(+\infty)\cr c_2(+\infty) \cr \vdots \cr c_n(+\infty)}.
\ee
Such a relation makes sense because of the linearity of the equation
(\ref{eqco}).
It is a well known result that under our general hypotheses, the
$S$-matrix
satisfies
\be
  S=\un +\ode (\eps).
\ee
Note that the $j^{\mbox{\scriptsize th}}$ column of the $S$-matrix
is given by the solution of (\ref{eqco}) at $t=\infty$
subjected to the initial
conditions $c_k(-\infty)=\delta_{jk}$, $k=1,2,\cdots ,n$.

In general, the $S$-matrix defined above has no particular properties
besides
that of being invertible. However, when the generator $H(t)$ satisfies
some
symmetry properties, the same is true for $S$. As such properties are
important in applications, we show below that if $H(t)$ is self-adjoint
 with respect to some indefinite scalar product, then
$S$ is unitary with respect to another indefinite scalar product.
Let $J\in M_n(\C)$ be an invertible self-adjoint matrix.
We define an indefinite metric on $\C^n$ by means of the indefinite scalar
product
\be
  (\cdot ,\cdot )_J=\bra \cdot |J\cdot \ket.
\ee
It is easy to check that the adjoint $A^{\#}$ of a matrix $A$ with
respect to the
$(\cdot ,\cdot )_J$ scalar product is given by
\be
  A^{\#}=J^{-1}A^*J.
\ee
\begin{prop}\label{PROS}
Let $H(t)$ satisfy H1, H2 and possess $n$ distinct eigenvalues $\forall
t\in\R$.
Further assume $H(t)$ is self-adjoint with respect to the
scalar product $(\cdot ,\cdot )_J$,
\be
  H(t)=H^{\#}(t)=J^{-1}H^*(t)J,\;\;\; \forall t\in \R,
\ee
and the eigenvectors $\ffi_j(0)$ of $H(0)$ satisfy
\be
  (\ffi_j(0) ,\ffi_j(0) )_J=\rho_j,\;\;\;\rho_j\in\{-1,1\},\;\;\;\forall
  j=1,\cdots, n.
\ee
Then the eigenvalues of $H(t)$ are real $\forall t\in\R$ and the
$S$-matrix is
unitary with respect to the scalar product
$(\cdot ,\cdot )_R$, where $R=R^*=R^{-1}$ is the real diagonal matrix
$R=\mbox{diag }(\rho_1,\rho_2,\cdots ,\rho_n)$,
\be
  S^{\#}=RS^*R=S^{-1}.
\ee
\end{prop}
{\bf Remark:}\\
The condition $(\ffi_j(0) ,\ffi_j(0) )_J=\pm 1$ can always be satisfied by
suitable renormalization provided $(\ffi_j(0) ,\ffi_j(0) )_J\neq 0$.

The main interest of this proposition is that when the $S$-matrix
possesses
symmetries, some of its elements can be deduced from
resulting identities, without resorting to their actual computations.

A simple proof of the proposition making use of notions discussed in the
next section
can be found in appendix.
The above proposition can
actually be used for the two main applications we deal with in section
\ref{applic}.
Note that in specific cases, further symmetry property can be derived for
the $S$-matrix,
see section \ref{applic}.

\section{Analyticity properties}\label{anprop}
\setcounter{equation}{0}
The generator $H(z)$ is analytic in $S_{\alpha}$, hence the solution of
the
linear equation (\ref{schr}) $\psi(z)$ is analytic in $S_{\alpha}$ as
well.
However, the eigenvalues and eigenprojectors of $H(z)$ may have
singularities in $S_{\alpha}$. Let us recall some basic
properties, the proofs of which can be found in \cite{k}. The eigenvalues
and eigenprojectors of a matrix analytic in a region of the complex plane
have analytic continuations in that region with
possible singularities located at points $z_0$ called exceptional points.
In a neighborhood free of exceptional points,
the eigenvalues are given by branches of analytic functions and their
multiplicities are constant.
One eigenvalue can therefore be analytically continued  until it
coincides at
$z_0$ with one or several other eigenvalues. The set of such points
defines the set of exceptional points.
The eigenvalues may possess branching points at an
exceptional $z_0$, where they are continuous, whereas the eigenprojectors
are
also multivalued but diverge as $z\ra z_0$. Hence, by hypothesis H3,
the $n$ distinct eigenvalues $e_j(t)$
defined on the real axis are analytic on the real axis and possess
multivalued analytic
continuations in $S_{\alpha}$, with possible branching points at the set
of
degeneracies $\Omega$, given by
\be
  \Omega=\left\{ z_0 |\; e_j(z_0)=e_k(z_0), \mbox{for some $k,j$ and
  some
  analytic continuation} \right\}.
\ee
By assumption H2, $\Omega$ is finite, by H3, $\Omega\cap\R=\emptyset$ and
$\Omega=\overline{\Omega}$, due to Schwarz's principle. Similarly, the
eigenprojectors $P_j(t)$ defined on the real axis are analytic on the real
axis and possess multivalued analytic continuations in $S_{\alpha}$, with
possible singularities at $\Omega$.
To see more precisely what happens to these multivalued functions
when we turn around a point $z_0\in\Omega$, we consider the
construction described in figure \ref{fig1}. Let $f$ be a multivalued
analytic function in
$S_{\alpha}\backslash\Omega$. We denote by $f(z)$ the analytic
continuation
of $f(0)$ along some path $\beta\in S_{\alpha}\backslash\Omega$ from $0$
to $z$.
Then we perform the analytic continuation of $f(z)$ along a negatively
oriented loop $\delta$ based at $z$ around a unique point $z_0\in\Omega$,
and denote by $\widetilde{f}(z)$ the function we get when we come back at
the starting point (if $\delta$ is positively oriented,
the construction is similar). For later purposes, we define $\eta_0$ as
the
negatively
oriented loop homotopic to the loop based at the origin encircling $z_0$
obtained by following
$\beta$ from $0$ to $z$, $\delta$ from $z$ back to $z$ and $\beta$ in the
reverse sense from $z$ back to the origin.
\begin{figure}
\vspace{5cm}
\hspace{3cm}\special{picture fig1 scaled 700}
\caption{The paths $\beta$, $\delta$ and $\eta_0$ in
$S_{\alpha}\backslash \Omega$.}\label{fig1}
\end{figure}
We'll keep this notation in the rest of this section.
It follows from the foregoing that if we perform the analytic continuation
of the set of eigenvalues $\{e_j(z)\}_{j=1}^n$, along a negatively
oriented loop around $z_0\in\Omega$, we get the set
$\{\widetilde{e}_j(z)\}_{j=1}^n$ with
\be
  \widetilde{e}_j(z)=e_{\sigma_0(j)}(z), \;\;j=1,\cdots,n,
\ee
where
\be
  \sigma_0 :\;\{1,2,\cdots,n\}\ra \{1,2,\cdots,n\}
\ee
is a permutation which depends on $\eta_0$. Similarly, and with the same
notations, we get for the analytic continuations of the projectors around
$z_0$
\be\label{prop}
  \widetilde{P}_j(z)=P_{\sigma_0(j)}(z), \;\;j=1,\cdots,n.
\ee
Let us consider now the eigenvectors $\ffi_j(t)$. We define $W(t)$ as the
solution of
\bea\label{par}
  W'(t)&=&\sum_{j=1}^nP_j'(t)P_j(t)W(t)\\
    &\equiv&K(t)W(t),\;\;\; W(0)=\un,\nonumber
\eea
where $t\in\R$. It is well known \cite{k}, \cite{kr}, that $W(t)$
satisfies the
intertwining identity
\be\label{inter}
  W(t)P_j(0)=P_j(t)W(t),\;\;\;j=1,2,\cdots,n\; ,\;\forall t\in\R,
\ee
so that, if
$\{\ffi_j(0)\}_{j=1}^n$
denotes a set of eigenvectors of $H(0)$, the vectors defined by
\be\label{defeig}
  \ffi_j(t)=W(t)\ffi_j(0)
\ee
are eigenvectors of $H(t)$. Moreover, using the identity
$Q(t)Q'(t)Q(t)\equiv 0$ which is true for any differentiable projector,
it
is easily checked that condition (\ref{phco}) is satisfied by these
vectors. The generator $K(t)$ is analytic on the real axis and can be
analytically continued in $S_{\alpha}\backslash \Omega$. Actually, $K(z)$
is single valued in $S_{\alpha}\backslash \Omega$. Indeed, let us consider
the analytic continuation of $K(z)$ around
$z_0\in\Omega$. We get from (\ref{prop}) that
\be
  \widetilde{P}_j'(z)=P_{\sigma_0(j)}'(z),
\ee
so that
\bea
  \widetilde{K}(z)&=&\sum_{j=1}^n\widetilde{P}_j'(z)\widetilde{P}_j(z)
  =\sum_{j=1}^n{P}_{\sigma_0(j)}'(z){P}_{\sigma_0(j)}(z)
  \nonumber\\
  &=&\sum_{k=1}^n{P}_{k}'(z){P}_{k}(z)=K(z).
\eea
Consequently, $W(t)$ can be analytically continued in
$S_{\alpha}\backslash \Omega$, where it is multivalued and satisfies
both
(\ref{par}) and (\ref{inter}) with $z\in S_{\alpha}\backslash \Omega$ in
 place
of $t\in\R$. Moreover, the relation between the analytic continuation
$W(z)$
from $0$ to some point $z\in S_{\alpha}\backslash \Omega$
and the analytic continuation $\widetilde{W}(z)$ is given by a monodromy
matrix $W(\eta_0)$ such that
\be\label{mono}
  \widetilde{W}(z)=W(z)W(\eta_0),
\ee
where $\eta_0$ is the negatively oriented loop based at the origin
which encircles $z_0\in\Omega$ only, (see figure
\ref{fig1}).
Note also that the analytic continuation $W(z)$ is invertible in
$S_{\alpha}\backslash \Omega$ and $W^{-1}(z)$ satisfies
\be
  {W^{-1}}'(z)=-W^{-1}(z)K(z),\;\;\; W^{-1}(0)=\un.
\ee
As a consequence, the eigenvectors (\ref{defeig}) possess multivalued
analytic extensions
in $S_{\alpha}\backslash \Omega$. Consider the relation
\be
   H(z)\ffi_j(z)=e_j(z)\ffi_j(z), \;\;
\ee
obtained by analytic continuation from $0$ to some point
$z\in S_{\alpha}\backslash \Omega$. When analytically
continued along a negatively oriented loop around $z_0\in\Omega$, it
yields
\be
  \widetilde{H}(z)\widetilde{\ffi}_j(z)=H(z)\widetilde{\ffi}_j(z)=
  \widetilde{e}_j(z)\widetilde{\ffi}_j(z)=e_{\sigma_0(j)}(z)
  \widetilde{\ffi}_j(z).
\ee
Thus $\widetilde{\ffi}_j(z)$ is proportional to ${\ffi}_{\sigma_0(j)}(z)$
and
we introduce the quantity $\theta_j(\eta_0)\in\C$ by the definition
\be\label{theta}
  \widetilde{\ffi}_j(z)=\e^{-i\theta_j(\eta_0)}{\ffi}_{\sigma_0(j)}(z),
  \;\;\;j=1,2,\cdots,n\;.
\ee
This is equivalent to (see (\ref{mono}))
\be
  W(\eta_0)\widetilde{\ffi}_j(0)=\e^{-i\theta_j(\eta_0)}
  {\ffi}_{\sigma_0(j)}(0).
\ee
Let us consider the couplings (\ref{coup}). Using the definition
(\ref{defeig}),
the invertibility of $W(t)$ and the identity (\ref{inter}), it's not
difficult
to see that we can rewrite
\be\label{anajk}
  a_{jk}(t)=-\frac{\bra\ffi_j(0)|P_j(0)W(t)^{-1}K(t)W(t)\ffi_k(0)\ket}
  {\|\ffi_j(0)\|^2},\;\;\; t\in\R,
\ee
which is analytic on the real axis and can be analytically continued in
$S_{\alpha}\backslash \Omega$, where it is multivalued. Thus, the same
is true
for the coefficients $c_j(t)$ which satisfy the linear differential
equation
(\ref{eqco}) and their analytic continuations satisfy the same equation
with
$z\in S_{\alpha}\backslash \Omega$ in place of $t\in \R$. We now come to
the
main identity of this section, regarding the coefficients $c_j(z)$. Let
us denote
by $c_j(z)$ the analytic continuation of $c_j(0)$ from $0$ to some
$z\in S_{\alpha}\backslash \Omega$. We perform
the analytic continuation of $c_j(z)$ along a negatively oriented loop
around
$z_0\in\Omega$ and denote by $\widetilde{c}_j(z)$ the function we get
when
we come back at the starting point $z$.
\begin{lem}\label{ECHCO}
For any $j=1,\cdots,n$, we have
\be\label{echco}
  \widetilde{c}_j(z)\e^{-i\int_{\eta_0}e_j(u)du/\eps}
  \e^{-i\theta_j(\eta_0)}
  =c_{\sigma_0(j)}(z)
\ee
where $\eta_0$, $\theta_j(\eta_0)$ and $\sigma_0(j)$ are defined as above.
\end{lem}
{\bf Proof:}\\
It follows from hypothesis H1 that $\psi(z)$ is analytic in $S_{\alpha}$
so that
\bea
  & &\sum_{j=1}^nc_j(z)\e^{-i\int_0^ze_j(u)du/\eps }\ffi_j(z)=
  \sum_{j=1}^n\widetilde{c}_j(z)\widetilde{\e^{-i\int_0^ze_j(u)du/\eps }}
  \widetilde{\ffi}_j(z)=\nonumber\\
  & &\sum_{j=1}^n\widetilde{c}_j(z){\e^{-i\int_{\eta_0}e_j(u)du/\eps }}
  \e^{-i\int_0^ze_{\sigma_0(j)}(u)du/\eps}\e^{-i\theta_j(\eta_0)}
  {\ffi}_{\sigma_0(j)}(z).
\eea
We conclude by the fact that $\{\ffi_j(z)\}_{j=1}^n$ is a basis.\ep \\
{\bf Remark:} \\
It is straightforward to generalize the study of the analytic
continuations around one singular point of the functions given above
to the case where the analytic continuations are performed around several
singular points, since $\Omega$ is finite. The loop $\eta_0$ can be
rewritten
as a finite succession of individual loops encircling one point of
$\Omega$
only, so that the permutation $\sigma_0$ is given by the composition of
a finite number of individual permutations. Thus
the factors $\e^{-i\theta_j(\eta_0)}$ in (\ref{theta}) should be replaced
by a
product of such factors, each associated with one individual loop and the
same
is true for the factors $\exp(-i\int_{\eta_0}e_j(z)dz/\eps )$ in lemma
\ref{ECHCO}. This process is performed in the proof of theorem
\ref{PERCO}.

\section{Complex WKB analysis}
\setcounter{equation}{0}
This section is devoted to basic estimates on the coefficients $c_j(z)$
in
certain domains extending to infinity in both the positive and negative
directions
inside the strip $S_{\alpha}$.
We first consider what happens in
neighborhoods of $\pm\infty$.
It follows from assumption H1 by a direct application of the Cauchy
formula
that
\be
 \lim_{t\ra\pm\infty}\sup_{|s|\leq\alpha}|t|^{1+a}\|H'(t+is)\|<\infty.
\ee
Hence the same is true for the single valued matrix $K(z)$
\be\label{deck}
  \lim_{t\ra\pm\infty}\sup_{|s|\leq\alpha}|t|^{1+a}\|K(t+is)\|<\infty.
\ee
Let $0<T\in\R$ be such that
\be\label{deft}
  \min_{z\in\Omega}\mbox{Re }z>-T,\;\;\;\mbox{and }
  \max_{z\in\Omega}\mbox{Re }z<+T.
\ee
All quantities encountered so far are analytic in
$S_{\alpha}\cap\{z||\mbox{Re }z|>T\}$, and we denote by a
"$\widetilde{\hspace{.2cm}}$" any
analytic continuation in that set. As noticed earlier
\be
  \widetilde{W}'(z)=K(z)\widetilde{W}(z),\;\;\; z\in S_{\alpha}
  \cap\{z||\mbox{Re }z|>T\}
\ee
so that it follows from (\ref{deck}) that the limits
\be
  \lim_{t\ra\pm\infty}\widetilde{W}(t+is)=\widetilde{W}(\pm\infty)
\ee
exist uniformly in $s\in ]-\alpha,\alpha[$.
Consequently, see (\ref{anajk}),
\be\label{refaj}
  \lim_{t\ra\pm\infty}|t|^{1+a}\sup_{|s|\leq\alpha}|\widetilde{a}_{jk}
  (t+is)|<\infty,
  \;\;\;\forall j,k\in\{1,\cdots,n\}.
\ee
Finally, for $|t|>T$, we can write
\bea
  \mbox{Im }\widetilde{\Delta}_{jk}(t+is)&=&\mbox{Im }\left(
  \int_{\eta}e_j(z)dz
  -\int_{\eta}e_k(z)dz\right)\nonumber\\
  &+&\int_0^s\mbox{Re }(e_{\sigma_j(j)}(t+is')-
  e_{\sigma_j(k)}(t+is'))ds',
\eea
where this equation is obtained by deforming the path of integration
from $0$ to
$z=t+is$ into a loop $\eta$ based at the origin, which may encircle
points of
$\Omega$, followed by the
real axis from $0$ to $\mbox{Re }z$ and a vertical path from
$\mbox{Re }z$
to $z$, see figure \ref{figlo}.
\begin{figure}
\vspace{4cm}
\hspace{3cm}\special{picture fig2 scaled 700}
\caption{The path of integration for $\widetilde{\Delta}_{jk}(z)$
(the x's denote points of $\Omega$).}\label{figlo}
\end{figure}
Hence we have
\be
  \sup_{z\in S_{\alpha}\cap\{z||\mbox{\scriptsize Re }z|>T\}}
  \mbox{Im }\widetilde{\Delta}_{jk}(z)<\infty,
\ee
which, together with (\ref{refaj})
yields the existence of the limits
\be\label{unili}
  \lim_{t\ra\pm\infty}\widetilde{c}_j(t+is)=\widetilde{c}_j(\pm\infty)
\ee
uniformly in $s\in ]-\alpha,\alpha[$.
We now define the domains in which useful estimates can be obtained.
\\
{\bf Definition:} {\em
Let $j\in\{1,\cdots,n\}$ be fixed.
A {\em dissipative domain for the index $j$},
$D_j\subset S_{\alpha}\backslash \Omega$, is such that
\be
  \sup_{z\in D_j}\mbox{\em Re }z=\infty,\;\;\;\inf_{z\in D_j}
   \mbox{\em Re }z=-\infty,
\ee
and is defined by the property that for any $z\in D_j$ and any
$k\in \{1,\cdots,n\}$, there exists a path
$\gamma^k\subset D_j$ parameterized by $u\in ]-\infty,t]$
which links $-\infty$ to $z$
\be
  \lim_{u\ra -\infty}\mbox{\em Re }\gamma^k(u)=-\infty,\;\;\;\gamma^k(t)
  =z
\ee
with
\be
  \sup_{z\in D_j}\sup_{u\in ]-\infty,t]}\left|\frac{d}{du}
  \gamma^k(u)\right|<\infty
\ee
and satisfies the monotonicity condition
\be
  \mbox{\em Im }\widetilde{\Delta}_{jk}(\gamma^k(u))\;\;\mbox{is a non
  decreasing
  function of }u\in]-\infty, t].
\ee
Such a path is a {\em dissipative path for $\{jk\}$}.
Here $\widetilde{\Delta}_{jk}(z)$ is the analytic continuation of
\be
  {\Delta}_{jk}(t)=\int_0^t(e_j(t')-e_k(t'))dt',\;\;\; t\in\R,
\ee
in $D_j$ along a path $\beta$ described in figure \ref{fig2} going from
 $0$ to
$-T\in\R$ along the real axis and
then vertically up or down until it reaches $D_j$, where $T>0$ is chosen
as in
(\ref{deft}).
}\\
{\bf Remark:}\\
The finiteness of $\Omega$ insures the existence of such a path $\beta$.
\begin{figure}
\vspace{5cm}
\hspace{3cm}\special{picture fig3 scaled 700}
\caption{The path $\beta$ along which the analytic continuation of
${\Delta}_{jk}(t)$
in $D_j$ is taken.}\label{fig2}
\end{figure}

Let $\widetilde{c}_k(z)$, $k=1,2,\cdots,n$, $z\in D_j$, be the analytic
continuations
of $c_k(t)$ along the same path $\beta$ which are solutions of the
analytic continuation of
(\ref{eqco}) in $D_j$ along $\beta$
\be\label{aneqc}
  \widetilde{c}_k'(z)=\sum_{l=1}^n\widetilde{a}_{kl}(z)
  \e^{i\widetilde{\Delta}_{kl}(z)
  /\eps}\widetilde{c}_l(z).
\ee
We take as initial conditions in $D_j$
\be\label{coin}
  \lim_{\mbox{\scriptsize Re }z\ra -\infty}\widetilde{c}_k(z)=
  \lim_{t\ra -\infty}c_k(t)=\delta_{jk},
  \;\;\;k=1,\cdots,n.
\ee
and we define
\be\label{defx}
  x_k(z)=\widetilde{c}_k(z)\e^{i\widetilde{\Delta}_{jk}(z)/\eps},\;\;\;
  z\in D_j,
  \;k=1,\cdots,n.
\ee
\begin{lem}\label{wkb}
  In a dissipative domain for the index $j$ we get the estimates
\bea
  & &\sup_{z\in D_j}|x_j(z)-1|=\ode (\eps)\\
  & &\sup_{z\in D_j}|x_k(z)|=\ode (\eps),\;\;\;\forall k\neq j.
\eea
\end{lem}
{\bf Remark:}\\
The real axis is a dissipative domain for all indices. In this case we
 have
$\widetilde{c}_j(t)\equiv c_j(t)$. Hence we get from
the application of the lemma for all indices successively that
\be\label{sibp}
  S=\un+\ode (\eps).
\ee
The estimates we are looking for are then just a direct corollary.
\begin{prop}\label{eslf}
Assume there exists a dissipative domain $D_j$ for the index $j$.
Let $\eta_j$ be a loop based at the origin which encircles all the
degeneracies between the real axis and $D_j$ and let $\sigma_j$ be the
permutation of labels associated with $\eta_j$, in the spirit of the
remark ending the previous section. The loop $\eta_j$ is negatively,
respectively
positively, oriented if $D_j$ is above, respectively below, the real axis.
Then the solution of
(\ref{eqco}) subjected to the initial conditions $c_k(-\infty)=
\delta_{jk}$
satisfies
\bea
  c_{\sigma_j(j)}(+\infty)&=&\e^{-i\theta_j(\eta_j)}
  \e^{-i\int_{\eta_j}e_j(z)dz/\eps}\left(1+\ode (\eps)\right)\\
  c_{\sigma_j(k)}(+\infty)&=&\ode \left(\eps\e^{
  \mbox{\em\scriptsize Im }\int_{\eta_j}e_j(z)dz/\eps+h_j
  (e_{\sigma_j(j)}(+\infty)-
  e_{\sigma_j(k)}(+\infty))/\eps}
  \right),
\eea
with $h_j\in [H^-_j,H^+_j]$, where $H^{\pm}_j$ is the maximum,
 respectively
minimum, imaginary part of the points at $+\infty$ in $D_j$
\be
  H^{+}=\lim_{t\ra +\infty}\sup_{s|t+is\in D_j}s,\;\;\;
  H^{-}=\lim_{t\ra +\infty}\inf_{s|t+is\in D_j}s.
\ee
\end{prop}
Thus we see that it is possible to get the (exponentially small)
asymptotic
behavior of the element $s_{\sigma_j(j),j}$ of the $S$-matrix, provided
there exists a dissipative domain for the index $j$. The difficult part
of the
problem is of course to prove the existence of such domains $D_j$, which
do not
necessarily exist, and to have enough of them to compute the asymptotic
of
the whole $S$-matrix.
This task is the equivalent for $n$-level systems to the study of the
global
behavior of the Stokes lines for $2$-level systems. We postpone this
aspect
of the problem to the next section. Note that we also get from this
result an
exponential bound on the elements $s_{\sigma_j(k),j}$ of the $S$-matrix,
$k\neq j$, which may or may not be useful. If $\eta_j$ encircles
no point of $\Omega$, we cannot get the asymptotic behavior of
$s_{\sigma_j(j),j}$ but we only get the exponential bounds. Since our
main
concern is asymptotic behaviors, we call the corresponding dissipative
domain
trivial.
\\ {\bf Remark:}\\
In contrast with the $2$-level case, see \cite{jp3}, we have to work
with dissipative
domains instead of working with one dissipative path for all indices.
Indeed, it is not difficult to convince oneself
with specific $3$-level cases that such a dissipative path may not
exist, even
when the eigenvalue degeneracies are close to the real axis. In return,
we prove below
the existence of dissipative domains in this situation.
\\
{\bf Proof:}\\
The asymptotic relation is a direct consequence of lemma \ref{ECHCO},
(\ref{unili}), (\ref{defx}) and the first part of the lemma. The estimate
is a
consequence of the same equations, the second estimate of the lemma and
the
identity, for $t>T$,
\bea
  \mbox{Im }\widetilde{\Delta}_{jk}(t+is)&=&\mbox{Im }\left(\int_{\eta_j}
  e_j(z)dz
  -\int_{\eta_j}e_k(z)dz\right)\nonumber\\
  &+&\int_0^s\mbox{Re }(e_{\sigma_j(j)}(t+is')-
  e_{\sigma_j(k)}(t+is'))ds'.
\eea
The path of integration from $0$ to $z$ for $\widetilde{\Delta}_{jk}(z)$
is deformed
into the loop $\eta_j$ followed by the
real axis from $0$ to $\mbox{Re }z$ and a vertical path from
$\mbox{Re }z$
to $z$.
It remains to take the limit $t\ra +\infty$.\ep\\
{\bf Proof of lemma \ref{wkb}:}\\
We rewrite equations (\ref{aneqc}) and (\ref{coin}) as an integral
equation and
perform an integration by parts on the exponentials
\bea
  \widetilde{c}_k(z)&=&\delta_{jk}-i\eps\sum_{l=1}^n
  \frac{\widetilde{a}_{kl}(z)}{\widetilde{e}_k(z)-\widetilde{e}_l(z)}
  \e^{i\widetilde{\Delta}_{kl}(z)/\eps}\widetilde{c}_l(z)\nonumber\\
  &+&i\eps\sum_{l=1}^n\int_{-\infty}^z
  {\left(\frac{\widetilde{a}_{kl}(z')}{\widetilde{e}_k(z')-
  \widetilde{e}_l(z')}\right)}'
  \e^{i\widetilde{\Delta}_{kl}(z')/\eps}\widetilde{c}_l(z')dz'\nonumber\\
  &+&i\eps\sum_{l,m=1}^n\int_{-\infty}^z
  \frac{\widetilde{a}_{kl}(z')\widetilde{a}_{lm}(z')}
  {\widetilde{e}_k(z')-\widetilde{e}_l(z')}
  \e^{i\widetilde{\Delta}_{km}(z')/\eps}\widetilde{c}_m(z')dz'.
\eea
Since all eigenvalues are distinct in $S_{\alpha}\backslash \Omega$, the
denominators are
always different from $0$. Due to equation (\ref{unili}), the height
above or below the
real axis
at which we start the integration is irrelevant, so that we can use the
symbol $-\infty$ as
lower integration bound. Note that the integrated term vanishes at
 $-\infty$. We have
also used the identity
\be
  \widetilde{\Delta}_{kl}(z')+\widetilde{\Delta}_{lm}(z')
  \equiv\widetilde{\Delta}_{km}(z').
\ee
In terms of the functions $x_k$ we get, using the same identity,
\bea\label{iefx}
  x_k(z)&=&\delta_{jk}-i\eps\sum_{l=1}^n
  \frac{\widetilde{a}_{kl}(z)}{\widetilde{e}_k(z)-\widetilde{e}_l(z)}
  x_l(z)\nonumber\\
  &+&i\eps\sum_{l=1}^n\int_{-\infty}^z
  {\left(\frac{\widetilde{a}_{kl}(z')}{\widetilde{e}_k(z')-
  \widetilde{e}_l(z')}\right)}'
  \e^{i(\widetilde{\Delta}_{jk}(z)-\widetilde{\Delta}_{jk}(z'))/\eps}
  x_l(z')dz'\nonumber\\
  &+&i\eps\sum_{l,m=1}^n\int_{-\infty}^z
  \frac{\widetilde{a}_{kl}(z')\widetilde{a}_{lm}(z')}{\widetilde{e}_k(z')
  -\widetilde{e}_l(z')}
  \e^{i(\widetilde{\Delta}_{jk}(z)-\widetilde{\Delta}_{jk}(z'))/\eps}
  x_m(z')dz'.
\eea
We introduce the quantity
\be
  |||x|||_j=\sup_{z\in D_j\atop l=1,\cdots,n}|x_l(z)|
\ee
and consider for each $k$ the equation (\ref{iefx}) along the dissipative
path $\gamma^k(u)$ described
in the definition of $D_j$, such that
\be
  \left|\e^{i(\widetilde{\Delta}_{jk}(\gamma^k(t))-
  \widetilde{\Delta}_{jk}(\gamma^k(u)))/\eps}\right|\leq 1
\ee
when $u\leq t$ along that path. Due to the integrability of the
$\widetilde{a}_{kl}(z)$ at infinity and the uniform boundedness of
$d\gamma^k(u)/du$, we get the estimate
\be
  |x_k(z)-\delta_{kj}|\leq \eps |||x|||_j A
\ee
for some constant $A$ uniform in $z\in D_j$, hence
\be
  |||x|||_j\leq 1+\eps |||x|||_j A.
\ee
Consequently, for $\eps$ small enough
\be
  |||x|||_j\leq 2.
\ee
And the result follows.\ep

\section{Superasymptotic improvement}\label{super}
\setcounter{equation}{0}

All results above can be substantially improved by using the so-called
superasymptotic renormalization method \cite{be}, \cite{n}, \cite{jp5}.
The joint use of complex WKB analysis
and superasymptotic renormalization is very powerful, as demonstrated
recently
in \cite{jp3} for $2$-level systems, and allows, roughly speaking, to
replace all
remainders
$\ode(\eps)$ by $\ode (\e^{-\kappa /\eps})$, where $\kappa >0$. We
 briefly
show how to achieve this improvement in the case of $n$-level systems.

Let $H(z)$ satisfy H1, H2 and H3 in $S_{\alpha}$ and let
\be
 \widehat{S}_{\alpha}=S_{\alpha}\backslash \cup_{r=1,\cdots,p}
 (J_r\cup\overline{J_r}),
\ee
where each $J_r$ is an open domain containing one point of $\Omega$ only
in the
open upper half plane. Hence, any analytic continuation $e_j(z)$ of
$e_j(t)$,
$t\in\R$, in
$\widehat{S}_{\alpha}$ is isolated in the spectrum of $H(z)$ so that
$e_j(z)$ is analytic and multivalued in $\widehat{S}_{\alpha}$, and the
same
is true for the corresponding analytic continuation $P_j(z)$ of
$P_j(t)$, $t\in\R$. Let $\sigma_r$ be the permutation associated with
the loop $\zeta_r$ based at the origin which encircles $J_r$ once, such
that
\be\label{mojr}
  \widetilde{e}_j(z)={e}_{\sigma_r(j)}(z),
\ee
with the convention of section \ref{anprop}. The matrix $K(z)$ is
analytic
and single valued in $\widehat{S}_{\alpha}$. Consider the single valued
analytic matrix
\be
  H_1(z,\eps)=H(z)-i\eps K(z),\;\;\; z\in \widehat{S}_{\alpha}.
\ee
For $\eps$ small enough, the spectrum of $H_1(z,\eps)$ is non degenerate
$\forall z\in \widehat{S}_{\alpha}$ so that its eigenvalues
$e_j^1(z,\eps)$ and
eigenprojectors $P_j^1(z,\eps)$ are multivalued analytic functions in
$\widehat{S}_{\alpha}$. Moreover, for $\eps$ small enough, the analytic
continuations of $e_j^1(z,\eps)$ and $P_j^1(z,\eps)$ around $J_r$
satisfy
\bea
  \widetilde{e}^1_j(z)&=&{e}^1_{\sigma_r(j)}(z)\\
  \widetilde{P}^1_j(z)&=&{P}^1_{\sigma_r(j)}(z),
\eea
as can be easily deduced from (\ref{mojr}) by perturbation theory.
 Consequently the
matrix
\be
  K_1(z,\eps)=\sum_{j=1}^m{{P}^1_j}'(z,\eps){P}^1_j(z,\eps)
\ee
is analytic and single valued in $\widehat{S}_{\alpha}$. Defining the
single
valued matrix
\be
  H_2(z,\eps)=H(z)-i\eps K_1(z,\eps),\;\;\; z\in \widehat{S}_{\alpha},
\ee
we can repeat the argument, for $\eps$ small enough. By induction we set
for
any $q\in\N$,
\bea
  H_q(z,\eps)&=&H(z)-i\eps K_{q-1}(z,\eps)\\
  K_{q-1}(z,\eps)&=&\sum_{j=1}^m{{P}^{q-1}_j}'(z,\eps){P}^{q-1}_j
  (z,\eps),
  \;\;\; z\in \widehat{S}_{\alpha}
\eea
for $\eps$ is small enough. We have
\be
  H_q(z,\eps)=\sum_{j=1}^m e^{q}_j(z,\eps){P}^{q}_j(z,\eps),
\ee
where the eigenvalues and eigenprojections are multivalued in
$\widehat{S}_{\alpha}$ and satisfy
\bea
  \widetilde{e}^{q}_j(z,\eps)&=&e^{q}_{\sigma_r(j)}(z,\eps)\\
  \widetilde{P}^{q}_j(z,\eps)&=&P^{q}_{\sigma_r(j)}(z,\eps),\;\;\; j=1,
  \cdots,n,
\eea
with the notations of (\ref{mojr}). We quote from \cite{jp3}, \cite{jp5}
the
main proposition regarding this construction.
\begin{prop}\label{supa}
  Let $H(z)$ satisfy H1, H2 and H3 in $S_{\alpha}$ and let
  $\widehat{S}_{\alpha}$ be defined as above. Then there exist constants
  $c>0$, $\eps^*>0$ and a real function $b(t)$ with
  $\lim_{t\ra\pm\infty}|t|^{1+a}b(t)<\infty$, such that
  \bea
    & &\|K_q(z,\eps)-K_{q-1}(z,\eps)\|\leq b(\mbox{\em Re }z)
    \eps^qc^qq!\\
    & &\|K_q(z,\eps)\|\leq b(\mbox{\em Re }z),
  \eea
  for all
  $z\in\widehat{S}_{\alpha}$, all $\eps<\eps^*$ and all
  $q\leq q^*(\eps)\equiv\left[1/ec\eps\right]$, where $[y]$ denotes the
  integer part of $y$ and $e$ is the basis of the neperian logarithm.
\end{prop}
We can deduce from this that in $\widehat{S}_{\alpha}$
\bea\label{peq}
  e_j^q(z,\eps)&=&e_j(z)+\ode (\eps^2b(\mbox{Re }z))\\
  P_j^q(z,\eps)&=&e_j(z)+\ode (\eps(\mbox{Re }z)),\;\;\;\forall
  q\leq q^*(\eps).
\eea
We introduce the notation $f^{q^*(\eps)}\equiv f^*$ for any quantity
$f^q$ depending on the index $q$ and we drop from now on the $\eps$ in
the arguments of the functions we encounter. We define the multivalued
analytic matrix $W_*(z)$ for $z\in \widehat{S}_{\alpha}$ by
\be
  {W_*}'(z)=K_*(z)W_*(z), \;\;\; W_*(0)=\un.
\ee
Due to the above observations and proposition \ref{supa}, $W_*(z)$ enjoys
all properties $W(z)$ does, such as
\bea
  & &W_*(z)P_j^*(z)=P_j^*(0)W_*(z)\\
  & &\widetilde{W}^*(z)=W_*(z)W_*(\zeta_r)
\eea
and, uniformly in $s$,
\be
  \lim_{t\pm\infty}W_*(t+is)=W_*(\infty).
\ee
Thus we define for any $z\in \widehat{S}_{\alpha}$ as set of
 eigenvectors of
$H_*(z)$ by
\be
  \ffi_j^*(z)= W_*(z) \ffi_j^*(0),
\ee
where
\be
   H_*(0)\ffi_j^*(0)=e^*_j(0)\ffi_j^*(0), \;\;\; j=1,\cdots, n,
\ee
and which satisfy
\be
  \widetilde{\ffi}_j^*(0)=\e^{-i\theta_j^*(\zeta_r)}
  \ffi_{\sigma_r(j)}^*(0),
\ee
with
\be
  \theta_j^*(\zeta_r)=\theta(\zeta_r)+\ode (\eps)\in \C.
\ee
Let us expand the solution of (\ref{schr}) on this multivalued set of
eigenvectors as
\be\label{decsup}
  \psi(z)=\sum_{j=1}^nc^*_j(z)\e^{-i\int_0^ze^*_j(z')dz'/\eps}\ffi^*_j(z).
\ee
Since the analyticity properties of the eigenvectors and eigenvalues of
$H_*(z)$ are
the same as those enjoyed by the eigenvectors and eigenvalues of $H(z)$,
we get
as in lemma \ref{ECHCO}
\be
  \widetilde{c}_j^*(z)\e^{-i\int_{\zeta_r}e_j^*(u)du/\eps}
  \e^{-i\theta_j^*(\zeta_r)}
  =c_{\sigma_r(j)}^*(z), \;\;\; \forall z\in \widehat{S}_{\alpha}.
\ee
Substituting (\ref{decsup}) in (\ref{schr}), we see that the multivalued
coefficients
${c}_j^*(z)$ satisfy in $\widehat{S}_{\alpha}$ the differential equation
\be
 {c_j^*}'(z)=\sum_{k=1}^na_{jk}^*(z)\e^{i\Delta_{jk}^*(z)/\eps}c_k^*(z)
\ee
where
\be
  \Delta_{jk}^*(z)=\int_0^ze_j^*(z')-e_k^*(z')dz'
\ee
and
\be
  a_{jk}^*(z)=\frac{\bra\ffi_j^*(z)(0)|P_j^*(z)(0){W_*(z)}^{-1}
  (K_{q^*-1}(z)-K_{q^*}(z))W_*(z)\ffi_k^*(0)\ket}{\|\ffi_j^*(0)\|^2},
\ee
to be compared with (\ref{anajk}).
The key point of this construction is that it follows from proposition
\ref{supa} with $q=q^*(\eps)$ that
\be\label{key}
  |a_{jk}^*(z)|\leq 2 b(\mbox{Re }z)\e^{-\kappa/\eps},\;\;\; \forall
  z\in \widehat{S}_{\alpha}
\ee
where $\kappa=1/ec >0$, and from (\ref{peq}) that
\be\label{keyd}
  \mbox{Im }\Delta_{jk}^*(z)=\mbox{Im }\Delta_{jk}(z)+\ode (\eps^2),
\ee
uniformly in $z\in \widehat{S}_{\alpha}$. Thus, we deduce from
(\ref{key})
that the limits
\be
  \lim_{t\ra\pm\infty}c_j^*(t+is)=c_j^*(\pm\infty), \;\;\; j=1,\cdots,n,
\ee
exist for any analytic continuation in $\widehat{S}_{\alpha}$. Moreover,
along any dissipative path $\gamma^k(u)$ for $\{jk\}$, as defined above,
we get from
(\ref{keyd})
\be
  \left|\e^{i(\widetilde{\Delta}_{jk}^*(\gamma^k(t))-
  \widetilde{\Delta}_{jk}^*(\gamma^k(u)))/\eps}\right|=\ode (1)
\ee
so that, reproducing the proof of lemma \ref{wkb} we have
\begin{lem}
  In a dissipative domain $D_j$, if
  $\widetilde{c}_k^*(-\infty)={c}_k^*(-\infty)=\delta_{kj}$, then
  \bea
    & &\widetilde{c}_j^*(z)=1+\ode (\e^{-\kappa/\eps})\\
    & &\e^{i\widetilde{\Delta}_{jk}(z)\eps}\widetilde{c}_k^*(z)=\ode
    \left(\e^{-\kappa/\eps}\right)
    ,\;\;\;\forall k\neq j,
  \eea
uniformly in $z\in \widehat{S}_{\alpha}$.
\end{lem}
This lemma
yields the improved version of our main result.
\begin{prop}\label{MAPROP}
Under the conditions of proposition \ref{eslf} and with the
same notations. If $c_k^*(-\infty)=\delta_{jk}$, then
\bea
  c_{\sigma_j(j)}^*(+\infty)&=&\e^{-i\theta_j^*(\eta_j)}
  \e^{-i\int_{\eta_j}e_j^*(z)dz/\eps}\left(1+\ode \left(\e^{-\kappa/\eps}
  \right)\right)\\
  c_{\sigma_j(k)}^*(+\infty)&=&\ode \left(\e^{-\kappa/\eps}\e^{
  \mbox{\em\scriptsize Im }\int_{\eta_j}e_j(z)dz/\eps+h_j(e_{\sigma_j(j)}
  (+\infty)-
  e_{\sigma_j(k)}(+\infty))/\eps}
  \right).
\eea
\end{prop}
Note that we may or may not replace $e_j(z)$ by $e_j^*(z)$ in the
estimate
without altering the result.
It remains to make the link between the $S$-matrix and the
$c_k^*(+\infty)$'s
of the proposition explicit. We define $\beta_j^{*\pm}$ by the
relations
($H_*(z)$ and $H(z)$ coincide at $\pm\infty$),
\be
  \ffi_j^*(\pm\infty)=\e^{-i\beta_j^{*\pm}}\ffi_j(\pm\infty).
\ee
By comparison of (\ref{decsup}) and (\ref{expa}) we deduce the lemma
\begin{lem}\label{sdec}
If $c_k(t)$ and $c_k^*(t)$ satisfy $c_k(-\infty)=c_k^*(-\infty)=
\delta_{jk}$,
  then, the element $kj$ of the $S$-matrix is given by
  \bea
    s_{kj}&=&c_k(+\infty)=\e^{-i(\beta_k^{*+}-\beta_j^{*-})}
    \e^{-i\int_0^{+\infty}
    e_k^*(t')-e_k(t')dt'/\eps}\e^{-i\int_{-\infty}^0e_j^*(t')-
    e_j(t')dt'/\eps}
    c_k^*(+\infty)\nonumber\\
    &\equiv& \e^{-i\alpha_{kj}^*}c_k^*(+\infty),
  \eea
  with $\beta_j^{*\pm}=\ode (\eps)$ and
  $\int_{\pm\infty}^0e_j^*(t')-e_j(t')dt'/\eps=\ode (\eps)$, i.e.
  $\e^{-i\alpha_{kj}^*}=1+\ode (\eps)$.
  \end{lem}
{\bf Remarks:}
\\ i) Proposition \ref{MAPROP} together with lemma \ref{sdec} are the
main results of the
first part of this paper.
\\ ii) As a direct consequence of these estimates on the real axis we
have
\be\label{sjksup}
  s_{jk}=\ode \left(\e^{-\kappa/\eps}\right),\;\;\; \forall k\neq j,
\ee
and
\be
  s_{jj}=\e^{-i\alpha_{jj}^*}\left(1+\ode \left(e^{-\kappa /\eps}\right)
  \right).
\ee
iii) It should be clear from the analysis just performed that all results
obtained hold if the generator $H(z)$ in (\ref{schr}) is replaced by
\be
  H(z,\eps)=H_0(z)+\ode (\eps b(\mbox{Re }z)),
\ee
with $b(t)=\ode (1/t^{1+a})$,
provided $H_0(z)$ satisfies the hypotheses we assumed.

\section{Avoided crossings}\label{avoid}
\setcounter{equation}{0}
We now come to the second part of the paper in which we prove asymptotic
formulas for
the off-diagonal elements of the $S$-matrix, by means of the general set
up presented
above.
To start with, we define a class of $n$-level systems for which we can
prove the
existence of one non trivial dissipative domains for all indices.
They are obtained by means
of systems exhibiting degeneracies of eigenvalues on the real axis,
hereafter
called real crossings, which we
perturb in such a way that these degeneracies are lifted and turn to
avoided
crossings on the real axis. When the perturbation is small enough, this
process
moves the eigenvalue degeneracies off the real axis but they
remain close to the place where the real crossings occurred.
This method was used successfully in \cite{j1} to deal with $2$-level
 systems.
We do not attempt to list
all cases in which dissipative domains can be constructed by means of
this technique
but rather present a wide class of examples which are relevant in the
theory of quantum adiabatic
transitions and in the theory of multichannel semiclassical scattering,
as described
below.
\\
Let $H(t,\delta)\in M_n(\C)$ satisfy the following assumptions.
\\
{\bf H4} {For each fixed $\delta\in[0,d]$, the matrix
$H(t,\delta)$ satisfies H1 in a strip $S_{\alpha}$ independent of
$\delta$ and $H(z,\delta)$,
$\partial /\partial z H(z,\delta)$ are
continuous as a functions of two variables $(z,\delta)\in S_{\alpha}
\times [0,d]$.
Moreover, it satisfies H2 uniformly in
$\delta\in[0,d]$, with limiting values $H(\pm,\delta)$ which are
continuous
functions of $\delta\in [0,d]$.}\\
{\bf H5} {\em For each $t\in\R$ and each $\delta\in[0,d]$, the spectrum
of $H(t,\delta)$, denoted
by $\sigma(t,\delta)$, consists in $n$ real eigenvalues
\be
  \sigma(t,\delta)=\{e_1(t,\delta), e_2(t,\delta),\cdots,e_n(t,\delta)\}
  \subset \R
\ee
which are distinct when $\delta>0$
\be
  e_1(t,\delta)< e_2(t,\delta)<\cdots <e_n(t,\delta).
\ee
When $\delta =0$, the functions $e_j(t,0)$ are analytic on the real axis
and
there exists a finite set
of crossing points $\{t_1\leq t_2\leq \cdots \leq t_p\}\in\R$, $p\geq 0$,
such that
\\
i) $\forall t<t_1$,
\be
  e_1(t,0)< e_2(t,0)<\cdots <e_n(t,0).
\ee
ii) $\forall j<k\in\{1,2,\cdots ,n\}$, there exists at most one $t_r$
with
\be
  e_j(t_r,0)-e_k(t_r,0)=0,
\ee
 and if such a $t_r$ exists we have
 \be
   \frac{\partial}{\partial t}\left(e_j(t_r,0)-e_k(t_r,0)\right)>0.
 \ee
 iii) $\forall j\in\{1,2,\cdots ,n\}$, the eigenvalue $e_j(t,0)$ crosses
 eigenvalues whose
 indices are all superior to $j$ or all inferior to $j$.
}

\noindent{\bf Remarks:}
\\ i) The parameter $\delta$ can be understood as a coupling constant
controlling
the strength of the perturbation.
\\ ii)The eigenvalues $e_j(t,0)$ are assumed to be analytic on the real
axis,
because of the degeneracies on the real axis. However, if $H(t,\delta)$
is self adjoint
for any $\delta\in[0,d]$, this follows from a theorem of Rellich, see
\cite{k}.
\\ iii) We give in figure \ref{figlev} an example of pattern of crossings
with
the corresponding pattern of avoided crossings for which the above
conditions are fulfilled.
\\ iv) The crossings are assumed to be generic
in the sense that the derivative of $e_j-e_k$ are non zero at the
crossing $t_r$.
\\ v) The crossing points $\{t_1,t_2,\cdots ,t_p\}$ need not be distinct,
which is important when
the eigenvalues possess symmetries. However, for each $j=1,\cdots, n$,
the eigenvalue
$e_j(t,\delta)$ experiences avoided crossings with $e_{j+1}(t,\delta)$
and/or
$e_{j-1}(t,\delta)$ at a subset of distinct points
$\{t_{r_1},\cdots ,t_{r_j}\}\subseteq \{t_1,t_2,\cdots ,t_p\}$.
\begin{figure}
\vspace{7cm}
\hspace{3cm}\special{picture fig4 scaled 500}
\caption{A pattern of eigenvalue crossings (bold curves) with
the corresponding pattern of avoided crossings (fine curves)
satisfying H5.}\label{figlev}
\end{figure}

We now state the main lemma of this section regarding the analyticity
properties of the perturbed levels and the existence of
dissipative domains for all indices in this perturbative context.
\begin{lem}\label{MAINL}
Let $H(t,\delta)$ satisfy H4 and H5. We can choose $\alpha>0$ small
 enough so
that the following assertions are true for sufficiently small
$\delta>0$:
\\ i) Let $\{t_{r_1},\cdots ,t_{r_j}\}$ be the set of avoided crossing
points experienced by $e_j(t,\delta)$, $j=1,\cdots, n$.
For each $j$, there exists a
set of distinct domains $J_r\in S_{\alpha}$, where
$r\in \{{r_1},\cdots ,{r_j}\}$,
\be
  J_r=\{z=t+is|\,0\leq |t-t_r|<L,\, 0<g<s<\alpha'\},
\ee
with $L$ small enough, $\alpha'<\alpha$ and $g>0$ such that
$e_j(-\infty,\delta)$ can be analytically continued in
\be
  S^j_{\alpha}=S_{\alpha}\backslash \cup_{r=r_1,\cdots,r_j}
  (J_r\cup\overline{J_r}).
\ee
ii) Let $t_r$ be an avoided crossing point of $e_j(t,\delta)$ with
$e_{k}(t,\delta)$, $k=j\pm 1$.
Then the analytic continuation of $e_j(t_r,\delta)$ along a loop based
at $t_r\in\R$ which encircles $J_r$ once yields
$\widetilde{e}_j(t_r,\delta)$ back at $t_r$ with
\be
 \widetilde{e}_j(t_r,\delta)=e_{k}(t_r,\delta).
\ee
iii) For each $j=1,\cdots, n$,
there exists a dissipative domain $D_j$ above or below the real axis in
$S_{\alpha}\cap\{z=t+is|\,|s|\geq \alpha'\}$.
The permutation $\sigma_j$ associated with these
dissipative domains (see proposition \ref{eslf}) are all given by
$\sigma_j=\sigma$
where $\sigma$ is the permutation which maps the index
of the $k^{\scriptsize \mbox{th}}$ eigenvalue $e_j(\infty,0)$ numbered
from the lowest one on $k$, for all $k\in\{1,2,\cdots ,n\}$.
\end{lem}
{\bf Remarks:}
\\ i) In part ii) the same result is true along a loop encircling
$\overline{J_r}$.
\\ ii) The dissipative domains $D_j$ of part iii) are located above
(respectively below) all the sets $J_r$ (resp. $\overline{J_r}$),
$r=1,\cdots, p$.
\\ iii) The main interest of this lemma is that the sufficient
conditions
required for the existence of dissipative domains in the complex plane
can be deduced from the behavior of the eigenvalues on the {\em real}
axis.
\\ iv) We emphasize that more general types of avoided crossings than
those described in
H5 may lead to the existence of dissipative domains for {\em certain}
indices
but we want to get dissipative domains for {\em all} indices. For
example, if
part iii) of H5 is satisfied for certain indices only, then part iii) of
lemma \ref{MAINL}
is satisfied for those indices only.
\\ v) Note also that there are patterns of eigenvalue crossings for which
there
exist no dissipative domain for some indices. For example, if $e_j(t,0)$
and $e_k(t,0)$
display two crossings, it is not difficult to see from the proof of the
lemma
that no dissipative domains can exist for $j$ or $k$.

We postpone the proof of this lemma to the end of the section and go on
with
its consequences.
By applying the results of the previous section we get
\begin{thm}\label{PERCO}
  Let $H(t,\delta)$ satisfy H4 and H5. If $\delta>0$ is small enough,
  the
  elements $\sigma(j)j$ of the $S$-matrix, with $\sigma(j)$ defined in
  lemma \ref{MAINL}, are given in the limit $\eps\ra 0$ for all $j=1,
  \cdots, n$, by
  \be
    s_{\sigma(j)j}=\prod_{k=j}^{\sigma(j)\mp1}\e^{-i\theta_k(\zeta_k)}
    \e^{-i\int_{\zeta_k}
    e_k(z,\delta)dz/\eps}\left(1+\ode (\eps)\right),\;\;\;\sigma(j)
    \left\{{ >j \atop <j}\right.
  \ee
  where, for $\sigma(j)>j$ (respectively $\sigma(j)<j$), $\zeta_k$,
  $k=j,\cdots,\sigma(j)\mp 1$, denotes a
  negatively (resp. positively) oriented loop
  based at the origin which encircles the set $J_r$ (resp.
  $\overline{J_r}$
  corresponding to the avoided crossing between $e_k(t,\delta)$ and
  $e_{k+1}(t,\delta)$ (resp. $e_{k-1}(z,\delta)$) at $t_r$,
  $\int_{\zeta_k}e_k(z,\delta)dz$ denotes the integral along $\zeta_k$
  of the
  analytic continuation of $e_k(0,\delta)$
  and $\theta_k(\zeta_k)$
  is the corresponding factor defined by (\ref{theta}), see figure
  \ref{ficor}.

  More accurately, with the notations of section \ref{super}, we have
  the improved formula
  \be\label{imfor}
    s_{\sigma(j)j}=\e^{-i\alpha_{\sigma(j)j}^*}
    \prod_{k=j}^{\sigma(j)\mp1}\e^{-i\theta_k^*(\zeta_k)}
    \e^{-i\int_{\zeta_k}
    e_k^*(z,\delta)dz/\eps}\left(1+\ode \left(e^{-\kappa /\eps}\right)
    \right),
    \;\;\;\sigma(j)\left\{{ >j \atop <j}\right.
  \ee

  The element $\sigma(l)j$, $l\neq j$, are estimated by
  \be
    s_{\sigma(l)j}=\ode \left(\eps \e^{h(e_{\sigma(j)}(\infty,\delta)-
    e_{\sigma(l)}(\infty,\delta))/\eps}\prod_{k=j}^{\sigma(j)\mp1}\e^{
    \mbox{\em \scriptsize Im }\int_{\zeta_k}
    e_k(z,\delta)dz/\eps} \right),\;\;\;\sigma(j)\left\{{ >j \atop <j}
    \right.
  \ee
  where $h$ is strictly positive (resp. negative) for $\sigma(j)>j$
  (respectively $\sigma(j)<j$).
\end{thm}
{\bf Remarks:}
\\ i) As the eigenvalues are continuous at the degeneracy points, we
have that
\be
  \lim_{\delta\ra 0}\e^{-i\int_{\zeta_k}e_k(z,\delta)dz}=0,\;\;\;\forall
   k=1,\cdots, p.
\ee
ii) The remainders $\ode(\eps)$ depend on $\delta$ but it should be
 possible to get
estimates which are valid as both $\eps$ and $\delta$ tend to zero, in
the spirit of
\cite{j1}, \cite{mn} and \cite{r}.
\\ iii) This result shows that one off-diagonal element per column of
the $S$-matrix at least can be
computed asymptotically. However, it is often possible to get more
elements by making use
of symmetries of the $S$-matrix.
Moreover, if there exist dissipative domains going
above or below other eigenvalue degeneracies further away in the complex
plane, other
elements of the $S$-matrix can be computed.
\\ iv) Finally, note that all starred quantities in (\ref{imfor})
depend on $\eps$.
\begin{figure}
\vspace{6cm}
\hspace{3cm}\special{picture fig5 scaled 600}
\caption{The loops $\eta_j$ and $\zeta_k$, $k=j,\cdots,\sigma(j)-1$.}
\label{ficor}
\end{figure}
\\{\bf Proof:} The first thing to determine is whether the loops
$\zeta_k$ are
above or below the real axis. Since the formulas we
deduce from the complex WKB analysis are asymptotic formulas, it suffices
to
choose the case which yields exponential decay of $s_{\sigma(j)j}$. It
is
readily checked in the proof of lemma \ref{MAINL} below that if
$\sigma(j)>j$,
$D_j$ is above the real axis and if $\sigma(j)<j$, $D_j$ is below the
 real axis.
Then it
remains to explain how to pass from the loop $\eta_j$ given in
proposition \ref{eslf} to the set of loops $\zeta_k$, $k=j,\cdots,
\sigma(j)-1$.
We briefly deal with the case $\sigma(j)>j$, the other case being
similar.
It follows from lemma \ref{MAINL} that we can deform $\eta_j$ into the
set of
loops $\zeta_k$, each associated with one avoided crossing, as described
in
figure \ref{ficor}. Thus we have
\be
  \int_{\eta_j}=\sum_{k=j}^{\sigma(j)-1}\int_{\zeta_k}
\ee
for the decay rate and, see \ref{mono},
\be
  W(\eta_j)=W(\zeta_{\sigma(j)-1})\cdots W(\zeta_{j+1})W(\zeta_j)
\ee
for the prefactors.
Let $\nu_j$ be a negatively oriented loop based at $t_r$
which encircles $J_r$ as described in lemma \ref{MAINL}. Consider now
the loop
$\zeta_j$ associated with
this avoided crossing and deform it to the path obtained by going from
$0$
to $t_r$ along the real axis, from $t_r$ to $t_r$ along $\nu_j$ and
back from $t_r$ to the origin along the real axis. By the point ii) of
the lemma
we get
\be
  \widetilde{e}_j(0,\delta)={e}_{j+1}(0,\delta)
\ee
along $\zeta_j$, and, accordingly (see (\ref{theta})),
\be
  \widetilde{\ffi}_j(0,\delta)=\e^{-i\theta_j(\zeta_j)}{\ffi}_{j+1}
  (0,\delta).
\ee
This justifies the first factor in the formula. By repeating the argument
at the next avoided crossings, keeping in mind that we get
${e}_{j+1}(0,\delta)$
at the end of $\zeta_j$ and so on, we get the final result.
The estimate on $s_{\sigma(l)j}$ is obtained by direct application of
lemma \ref{MAINL}.
\ep
\\
{\bf Proof of lemma \ref{MAINL}:}
In the sequel we shall denote "$\frac{\partial}{\partial t}$" by a
"$\prime$ ".
We have to consider the analyticity properties of
$\widetilde{e}_j(z,\delta)$
and define domains in which every point $z$ can be reached
from $-\infty$ by means of a path $\gamma(u)$, $u\in]-\infty,t]$,
$\gamma(t)=z$ such that
$\mbox{Im }\widetilde{\Delta}_{jk}(\gamma(u),\delta)$
is non decreasing in $u$ for certain indices $j\neq k$ when $\delta >0$
is fixed.
Note that by Schwarz's principle if
$\gamma(u)$ is dissipative for $\{jk\}$, then $\overline{\gamma(u)}$ is
dissipative for $\{kj\}$.
When $\gamma(u)=\gamma_1(u)+i\gamma_2(u)$ is differentiable, saying that
$\gamma(u)$
is dissipative for $\{jk\}$ is equivalent to
\bea\label{disco}
  & &\mbox{Re }(\widetilde{e}_j(\gamma(u),\delta)-
  \widetilde{e}_k(\gamma(u),\delta))\dot{\gamma}_2(u)
  +\mbox{Im }(\widetilde{e}_j(\gamma(u),\delta)-
  \widetilde{e}_k(\gamma(u),\delta))\dot{\gamma}_1(u)
  \geq 0\nonumber\\
  & &\hfill\forall u\in]-\infty,t]
\eea
where "$\dot{\hspace{.2cm}}$" denotes the derivative with respect to $u$.
Moreover, if the eigenvalues are analytic in a neighborhood of the real
axis,
we have the relation in that neighborhood
\be\label{imre}
  \mbox{Im }(\widetilde{e}_j(t+is,\delta)-
  \widetilde{e}_k(t+is,\delta))=\int_0^s\mbox{Re }
  (\widetilde{e}_j'(t+is',\delta)-
  \widetilde{e}_k'(t+is',\delta))ds',
\ee
which is a consequence of the Cauchy-Riemann identity.
We proceed as follows. We construct dissipative domains above and below
the real
axis when $\delta=0$ and we show that they remain dissipative for the
perturbed quantities $\widetilde{\Delta}_{jk}(z,\delta)$ provided
$\delta$ is small enough.
We introduce some quantities to be used in the construction. Let
$C_r\in \{ 1,\cdots,n\}^2$ denote
the set of distinct couples of indices such that the corresponding
eigenvalues experience
one crossing at $t=t_r$. Similarly, $N\in \{ 1,\cdots,n\}^2$ denotes
the set of couples of indices such that the corresponding eigenvalues
never
cross.
\\Let $I_r=[t_r-L,t_r+L]\in\R$, $r=1,\cdots,p$, with $L$ so small that
\be
  \min_{r\in \{ 1,\cdots,p\}}\min_{\{jk\}\in C_r,\, j<k}\inf_{t\in I_r}
  (e_j'(t,0)-e_{k}'(t,0))\equiv 4c>0.
\ee
This relation defines the constant $c$ and we also define $b$ by
\bea
  & &\min_{r\in \{ 1,\cdots,p\}}\min_{\{jk\}\in C_r,\, j<k}
  \inf_{t\in \R\backslash I_r}
  \left|e_j(t,0)-e_{k}(t,0)\right|\geq 4b>0\\
  & &\min_{\{jk\}\in N,\, j<k}\inf_{t\in \R}\left|e_j(t,0)-e_{k}(t,0)
  \right|\geq 4b>0.
\eea
We further introduce
\be
  I_r^{\alpha}=\{z=t+is|t\in I_r, |s|\leq \alpha\},\;\;\; r=1,\cdots, p.
\ee
Then we choose $\alpha$ small enough so that the only points of
degeneracy of eigenvalues
in $S_{\alpha}$ are on the real axis and
\bea
 & &\label{dec}
 \min_{r\in \{ 1,\cdots,p\}}\min_{\{jk\}\in C_r,\, j<k}
 \inf_{z\in I_r^{\alpha}}
 \mbox{Re }(e_j'(z,0)-e_{k}'(z,0))> 2c>0\\
 & &\min_{r\in \{ 1,\cdots,p\}}\min_{\{jk\}\in C_r,\, j<k}
 \inf_{z\in S_{\alpha}
 \backslash I_r^{\alpha}}\left|
 \mbox{Re }(e_j(z,0)-e_{k}(z,0))\right|> 2b>0\\
 & &\min_{\{jk\}\in N,\, j<k}\inf_{z\in S_{\alpha}}\left|
 \mbox{Re }(e_j(z,0)-e_{k}(z,0))\right|> 2b>0.
\eea
That this choice is always possible is a consequence of the analyticity
of $e_j(z,0)$ close to
the real axis and of the fact that we can work essentially in a compact,
because of hypothesis H4. Let $a(t)$ be integrable on $\R$ and such
that
\be\label{defa}
  \frac{a(t)}{2}>\max_{j< k\in\{ 1,\cdots,n\}}\sup_{|s|\leq\alpha}
  \left|\mbox{Re }
  (e_j'(t+is,0)-e_{k}'(t+is,0))\right|.
\ee
It follows from H4 that such functions exist.

Let $r\in\{1,\cdots,p\}$ and $\gamma_2(u)$ be a solution of
\be
  \left\{\matrix{\dot{\gamma}_2(u)=-\frac{\gamma_2(u)a(u)}{b}
  \hfill& u\in]-\infty,t_r-L]\hfill\cr
         \dot{\gamma}_2(u)=0 \hfill& u\in]t_r-L,t_r+L[\hfill\cr
         \dot{\gamma}_2(u)=+\frac{\gamma_2(u)a(u)}{b}\hfill& u
         \in[t_r+L,\infty[\hfill}\right.
\ee
with $\gamma_2(t_r)>0$. Then $\gamma_2(u)>0$ for any $u$, since
\be
  \left\{\matrix{{\gamma}_2(u)=\gamma_2(t_r)\e^{-\int_{t_r-L}^u
  a(u')du'/b}
  \hfill& u\in]-\infty,t_r-L]\hfill\cr
         {\gamma}_2(u)=\gamma_2(t_r) \hfill& u\in]t_r-L,t_r+L[
         \hfill\cr
         {\gamma}_2(u)=\gamma_2(t_r)\e^{\int_{t_r+L}^ua(u')du'/b}
         \hfill& u\in[t_r+L,\infty[\hfill}\right.
\ee
and since $a(u)$ is integrable, the limits
\be
  \lim_{u\ra\pm\infty}\gamma_2(u)=\gamma_2(\pm\infty)
\ee
exist. Moreover, we can always choose $\gamma_2(t_r)>0$ sufficiently
 small so that
$\gamma^r(u)\equiv u+i\gamma_2(u)\in S_{\alpha}$, for any real $u$.
 Let us verify
that this path is dissipative for all $\{jk\}\in C_r,\, j<k$.
For $u\in ]-\infty,t_r-L]$, we have, using
\be
  \mbox{Re }({e}_j(z,0)-{e}_k(z,0))< -2b<0,\;\;\;\forall
  z\in S_{\alpha}\cap
  \{z|\mbox{Re }z\leq t_r-L\}
\ee
and
\be
  \left|\mbox{Im }({e}_j(t+is,0)-{e}_k(t+is,0))\right|<
  |s|\sup_{s'\in[0,s]}
  \left|\mbox{Re }({e}_j'(t+is',0)-{e}_k'(t+is',0))\right|,
\ee
(see (\ref{imre})), and the definition (\ref{defa})
\bea\label{anco}
  & &\mbox{Re }({e}_j(\gamma^r(u),0)-
  {e}_k(\gamma^r(u),0))\dot{\gamma}_2(u)
  +\mbox{Im }({e}_j(\gamma^r(u),0)-
  {e}_k(\gamma^r(u),0))\dot{\gamma}_1(u)
  =\nonumber\\
  & &-\mbox{Re }({e}_j(\gamma^r(u),0)-
  {e}_k(\gamma^r(u),0))\frac{\gamma_2(u)a(u)}{b}+
  \mbox{Im }({e}_j(\gamma^r(u),0)-
  {e}_k(\gamma^r(u),0))>\nonumber\\
  & &2\gamma_2(u)a(u)-\gamma_2(u)a(u)/2>\gamma_2(u)a(u)>0.
\eea
Similarly, when $u\geq t_r+L$ we get, using
\be
  \mbox{Re }({e}_j(z,0)-{e}_k(z,0))> 2b>0,\;\;\;\forall
  z\in S_{\alpha}\cap
  \{z|\mbox{Re }z\geq t_r+L\},
\ee
\bea\label{banco}
  & &\mbox{Re }({e}_j(\gamma^r(u),0)-
  {e}_k(\gamma^r(u),0))\dot{\gamma}_2(u)
  +\mbox{Im }({e}_1(\gamma^r(u),0)-
  {e}_k(\gamma^r(u),0))\dot{\gamma}_1(u)
  =\nonumber\\
  & &\mbox{Re }({e}_j(\gamma^r(u),0)-
  {e}_k(\gamma^r(u),0))\frac{\gamma_2(u)a(u)}{b}+
  \mbox{Im }({e}_j(\gamma^r(u),0)-
  {e}_k(\gamma^r(u),0))>\nonumber\\
  & &2\gamma_2(u)a(u)-\gamma_2(u)a(u)/2>\gamma_2(u)a(u)>0.
\eea
Finally, for $s\in [t_r-L,t_r+L]$, we have with (\ref{dec})
\bea\label{bancor}
  & &\mbox{Im }({e}_j(\gamma^r(u),0)-{e}_k(\gamma^r(u),0))=
  \int_0^{\gamma_2(u)}
  \mbox{Re }({e}_j'(t'+is,0)-{e}_k'(t'+is,0))\geq\nonumber\\
  & &\gamma_2(u)2c>\gamma_2(u)c>0.
\eea
Thus, $\gamma^r(u)$ is dissipative for all $\{jk\}\in C_r,\, j<k$.
Note that the last estimate shows that it is not possible to find a
dissipative
path for $\{jk\}\in C_r,\, j<k$ below the real axis.

Consider now $\{jk\}\in N,\, j<k$ and let $\gamma_2(u)$ be a solution of
\be
  \dot{\gamma}_2(u)=-\frac{\gamma_2(u)a(u)}{b}, \;\;\; \gamma_2(0)>0,
  \;u\in]-\infty,+\infty [,
\ee
i.e.
\be
  {\gamma}_2(u)=\gamma_2(0)\e^{-\int_{0}^ua(u')du'/b}.
\ee
As above, we have $\gamma_2(u)>0$ for any $u$ and we can choose
$\gamma_2(0)>0$ small enough so that
$\gamma^+(u)\equiv u+i\gamma_2(u)\in S_{\alpha}$ for any $u\in\R$.
Since
\be
  \mbox{Re }({e}_j(z,0)-{e}_k(z,0))> -2b\;\;\; \forall z\in S_{\alpha}
\ee
we check by a computation analogous to (\ref{anco}) that $\gamma^+(u)$
is
dissipative for $\{jk\}\in N,\, j<k$. Similarly, one verifies that if
$\gamma_2(u)$ is the solution of
\be
  \dot{\gamma}_2(u)=\frac{\gamma_2(u)a(u)}{b}, \;\;\; \gamma_2(0)<0,
  \;u\in]-\infty,+\infty [
\ee
with $|\gamma_2(0)|$ small enough, the path $\gamma^-(u)\equiv u+i
\gamma_2(u)$ below
the real axis is in $S_{\alpha}$ for any $u\in\R$ and is dissipative for
$\{jk\}\in N,\, j<k$ as well.

Finally, the complex conjugates of these paths yield dissipative paths
 above and
below the real axis for $\{jk\}\in N,\, j>k$.

We now define the dissipative domains by means of their
borders. Let $\gamma^+(u)$ and $\gamma^-(u)$, $u\in\R$, two dissipative
paths in
$S_{\alpha}$ defined as above with $|\gamma_2^-(0)|$ sufficiently small
so that $\overline{\gamma^-}$ is below $\gamma^+$. We set
\be
  D=\{z=t+is|0<-\gamma_2^-(t)\leq s\leq\gamma_2^+(t),\,t\in\R\}.
\ee
Let $z\in D$, and $j\in \{1,\cdots, n\}$ be fixed. By assumption H5, the
set
$X_j$ of indices $k$ such that
$\{jk\}\in C_r$ for some $r\in \{1,\cdots, p\}$ consists in values $k$
satisfying $j<k$ or
it consists in values $k$ satisfying $j>k$. Let us assume the first
alternative
takes place.
Now for any $k\in \{1,\cdots, n\}$, there are three cases.
\\ 1) If $k\in X_j$, then
there exists a dissipative path $\gamma^r\in D$ for $\{jk\}\in C_r,\,
 j<k$
constructed as above
which links $-\infty$ to $z$. It is enough to select the initial condition
$\gamma_2(t_r)$ suitably, see figure \ref{figdis}.
\\ 2) Similarly, if $j<k\not \in X_j$, there
exists a dissipative path $\gamma^+\in D$ for $\{jk\}$ constructed as
above
which links $-\infty$ to $z$ obtained by a suitable choice of
$\gamma_2(0)$.
\\ 3) Finally, if $k>j$, we can take as a dissipative path for $\{jk\}$,
the path
$\overline{\gamma^-}\in D$ constructed as above
which links $-\infty$ to $z$ with suitable choice of $\gamma_2(0)$.
Hence $D$ is
dissipative for the index $j$, when $\delta = 0$.
\begin{figure}
\vspace{5cm}
\hspace{3cm}\special{picture fig6 scaled 700}
\caption{The dissipative domain $D$ and some dissipative paths.}
\label{figdis}
\end{figure}
If $j$ is such that the set $X_j$ consists in points $k$ with $k>j$,
a similar argument
with the complex conjugates of the above paths shows that the domain
$\overline{D}$ below
the real axis is dissipative for $j$ when $\delta = 0$.

Let us show that these domains remain dissipative when $\delta>0$
is not too large. We start by considering the analyticity properties of
the perturbed eigenvalues $e_j(z,\delta)$, $\delta>0$.
Let $0<\alpha'<\alpha$ be such that
\be
  I_r^{\alpha'}\cap (D\cup \overline{D})=\emptyset\, ,\;\;\;\forall
  r=1,\cdots, p.
\ee
The analytic eigenvalues $e_j(z,0)$, $j\in \{1,\cdots ,n\}$, are
isolated in the spectrum of $H(z,0)$ for any
$z\in \widetilde{S}_{\alpha}$,
where
\be
  \widetilde{S}_{\alpha}={S}_{\alpha}\backslash
  \cup_{r=1,\cdots ,p}I_r^{\alpha'}.
\ee
For any $j=1,\cdots, n$ we get from perturbation theory \cite{k}, that
the analytic
continuations $\widetilde{e}_j(z,\delta)$ of $e_j(t_1-L,\delta)$ in
$\widetilde{S}_{\alpha}$ are all distinct in $\widetilde{S}_{\alpha}$,
provided $\delta$ is small enough. This is due to the fact that
assumption H5 implies the continuity of
$H(z,\delta)$ in $\delta$ uniformly in $z\in S_{\alpha}$, as
is easily verified.
More precisely, for any fixed index $j$, the eigenvalue $e_j(t,\delta)$
experiences avoided crossings at the points $\{t_{r_1},\cdots ,t_{r_j}\}$.
We can assume without loss of generality that
\be
  I_k^{\alpha'}\cap I_l^{\alpha'}=\emptyset \, , \;\;\; \forall k\neq l
  \in
  \{{r_1},\cdots ,{r_j}\}.
\ee
Hence, for $\delta>0$ small enough, the analytic continuation
$\widetilde{e}_j(z,\delta)$ is isolated in the spectrum of $H(z,\delta)$,
uniformly in $z\in S_{\alpha}\backslash \cup_{r={r_1},\cdots ,{r_j}}
I_r^{\alpha'}$. Since by assumption H5 there is no crossing of
eigenvalues
on the real axis when $\delta>0$, there exists a $0<g<\alpha'$, which
depends on $\delta$,
such that
$\widetilde{e}_j(z,\delta)$ is isolated in the spectrum of $H(z,\delta)$,
uniformly in $z\in {S}^j_{\alpha}$, where
\be
  {S}^j_{\alpha}={S}_{\alpha}\backslash \cup_{r={r_1},\cdots ,{r_j}}
  (J_r\cup\overline{J_r})
\ee
and
\be
  J_r=I_r^{\alpha'}\cap\{z| \,\mbox{Im }z>g\}\, , \;\;\; r=1,\cdots, p.
\ee
Hence the singularities of $\widetilde{e}_j(z,\delta)$ are located in
$\cup_{r={r_1},\cdots ,{r_j}}(J_r\cup\overline{J_r})$, which yields the
first assertion of the
lemma.

Consider a path $\nu_r$ from $t_r-L$ to $t_r+L$ which goes above $J_r$,
where $t_r$ is an avoided crossing between ${e}_j(t,\delta)$ and
${e}_{k}(t,\delta)$, $k=j\pm 1$.
By perturbation theory again, ${e}_j(t_r-L,\delta)$ and
${e}_k(t_r-L,\delta)$ tend to ${e}_{j'}(t_r-L,0)$ and ${e}_{k'}(t_r-L,0)$
as
$\delta\ra 0$, for some $j',k'\in 1,\cdots, n$, whereas ${e}_j(t_r+L,
\delta)$
and ${e}_k(t_r+L,\delta)$
tend to ${e}_{k'}(t_r+L,0)$ and ${e}_{j'}(t_r+L,0)$ as $\delta\ra 0$,
see figure \ref{figlev}. Now, the analytic continuations
of ${e}_j(t_r-L,\delta)$ and
${e}_k(t_r-L,\delta)$ along $\nu_r$,
$\widetilde{e}_j(z,\delta)$ and
$\widetilde{e}_k(z,\delta)$ tend to the analytic functions
$\widetilde{e}_{j'}(z,0)={e}_{j'}(z,0)$ and
$\widetilde{e}_{k'}(z,0)={e}_{k'}(z,0)$ as
$\delta\ra 0$, for all $z\in\nu_r$. Thus, we deduce that
for $\delta$ small enough
\be
  \widetilde{e}_j(t_r+L,\delta)\equiv {e}_k(t_r+L,\delta),
\ee
since we know that
$\widetilde{e}_j(t_r+L,\delta)={e}_{\sigma(j)}(t_r+L,\delta)$,
for some permutation $\sigma$.
Hence the point iii) of the lemma follows.

Note that the analytic
continuations $\widetilde{e}_j(z,\delta)$ are single valued in
$\widetilde{S}_{\alpha}$. Indeed, the analytic continuation of
${e}_j(t_r-L,\delta)$ along
$\overline{\nu_r}$, denoted by $\widehat{e}_j(z,\delta)$,
$\forall z\in \overline{\nu_r}$, is such that
\be
  \widehat{e}_j(t_r+L,\delta)=\overline{\widetilde{e}_j(t_r+L,\delta)}=
  \widetilde{e}_j(t_r+L,\delta)={e}_k(t_r+L,\delta),
\ee
due to Schwarz's principle.
We further require $\delta$ to be sufficiently small so that
the following estimates are satisfied
\bea
 & &\label{a}\min_{r\in \{ 1,\cdots,p\}}\min_{\{jk\}\in C_r \atop j<k}
 \inf_{z\in
 \widetilde{S}_{\alpha}\backslash I_r^{\alpha}}\left|
 \mbox{Re }(\widetilde{e}_j(z,\delta)-\widetilde{e}_{k}(z,\delta))
 \right|
 > b>0\\
 & &\label{be}\min_{\{jk\}\in N \atop j<k}\inf_{z\in
 \widetilde{S}_{\alpha}}\left|
 \mbox{Re }(\widetilde{e}_j(z,\delta)-\widetilde{e}_{k}(z,\delta))
 \right|
 >b>0\\
 & &\label{c}\max_{j<k\in \{ 1,\cdots,n\}}\sup_{\mbox{Im }z |\,z\in
 \widetilde{S}_{\alpha}}
 \left|\mbox{Re }
  (\widetilde{e}_j'(z,\delta)-\widetilde{e}_{k}'(z,\delta))\right|<
  a(\mbox{Re }z).
\eea
and, in the compacts $\widetilde{I}_r^{\alpha}={I}_r^{\alpha}\backslash
{I}_r^{\alpha'}$,
\bea
  & &\label{1}
  \min_{r\in \{ 1,\cdots,p\}}\min_{\{jk\}\in C_r \atop j<k}
  \inf_{z\in\widetilde{I}_r^{\alpha}}|\mbox{Im }
  (\widetilde{e}_j(z,\delta)-\widetilde{e}_{k}(z,\delta))|
  >\nonumber\\
  & &\frac{1}{2}\min_{r\in \{ 1,\cdots,p\}}\min_{\{jk\}\in C_r \atop j<k}
  \inf_{z\in\widetilde{I}_r^{\alpha}}|\mbox{Im }
  (\widetilde{e}_j(z,0)-\widetilde{e}_{k}(z,0))| > |\mbox{Im }z| c\\
  & &\label{2}
  \max_{r\in \{ 1,\cdots,p\}}\max_{j<k\in \{ 1,\cdots,n\}}\sup_{z\in
  \widetilde{I}_r^{\alpha}}\left|
  \mbox{Im }(\widetilde{e}_j(z,\delta)-\widetilde{e}_{k}(z,\delta))
  \right|<
  \nonumber\\
  & &2\max_{r\in \{ 1,\cdots,p\}}\max_{j<k\in \{ 1,\cdots,n\}}
  \sup_{z\in
  \widetilde{I}_r^{\alpha}}\left|\mbox{Im }
  (\widetilde{e}_j(z,0)-\widetilde{e}_{k}(z,0))\right|<
  |\mbox{Im }z| a(\mbox{Re }z).
\eea
The simultaneous requirements (\ref{defa}) and (\ref{c}) is made
possible by
the continuity properties of $H'(z,\delta)$ and the uniformity in
$\delta$
of the decay at $\pm\infty$ of $H(z,\delta)$ assumed in H4 together with
the fact that
$a(t)$ can be replaced by a multiple of $a(t)$ if necessary, to satisfy
both estimates.
The condition on $\delta$ is given by the first inequalities in (\ref{1})
and (\ref{2}),
whereas the second ones are just recalls.

Then it remains to check that the paths
$\gamma^r, \gamma^+$ and $\gamma^-$ defined above satisfy the
dissipativity
condition (\ref{disco}) for the corresponding indices.
This is not difficult,
since the above estimates are precisely designed to preserve inequalities
such as
(\ref{anco}), (\ref{banco}) and (\ref{bancor}). However, it should not
be forgotten that in the sets $I_r^{\alpha'}$ the eigenvalues may be
singular so that
(\ref{imre}) cannot be used there. So when checking that a path
parameterized
as above by $u\in \R$ is dissipative, it is necessary to consider
separately
the case $u\in \R\backslash (\cup_{r=1,\cdots,p}I_r)$, where we proceed
as above with (\ref{a}),
(\ref{be}), (\ref{c}) and (\ref{imre}) and the case
$u\in \cup_{r=1,\cdots,p}I_r$, where we use use (\ref{1}) and (\ref{2})
instead of (\ref{imre}) as follows.
If $u\in I_r$ for $r$ such that $t_r$ is a crossing point for
$e_j(t,0)$ and $e_k(t,0)$, one takes (\ref{1}) to estimate
$\mbox{Im }(\widetilde{e}_{j'}(z,\delta)-\widetilde{e}_{k'}(z,\delta))$
for the corresponding indices $j'$ and $k'$,
and if $t_r$ is not a crossing
point for $e_j(t,0)$ and $e_k(t,0)$, one uses (\ref{2}) to estimate
$\mbox{Im }(\widetilde{e}_{j'}(z,\delta)-\widetilde{e}_{k'}(z,\delta))$.
Consequently, the domains $D$ and
$\overline{D}$ defined above keep the same dissipativity properties when
$\delta>0$ is small enough.

Let us finally turn to the determination of the associated permutation
$\sigma$. As noticed earlier, the eigenvalues $\widetilde{e}_j(z,\delta)$
are continuous in $\delta$, uniformly in $z\in \widetilde{S}_{\alpha}$.
Hence,
since the eigenvalues ${e}_j(z,0)$ are analytic in $S_{\alpha}$, we have
\be
  \lim_{\delta\ra 0}\widetilde{e}_j(\infty,\delta)={e}_j(\infty,0)
  \;\;\; j=1,2,\cdots,n.
\ee
Whereas we have along the real axis (see figure \ref{figlev}),
\be
  \lim_{\delta\ra 0}{e}_{\sigma(j)}(\infty,\delta)={e}_j(\infty,0),
\ee
with $\sigma$ defined in the lemma,
from which the result follows.\ep

\section{Applications}\label{applic}
\setcounter{equation}{0}

Let us consider the time-dependent Schr\"odinger equation in the
adiabatic limit.
The relevant equation is then
(\ref{schr}) where  $H(t)=H^*(t)$ is the time-dependent self-adjoint
hamiltonian.
Thus we can take $J=\un$ in proposition \ref{PROS} to get
\be
  H(t)=H^*(t)=H^{\#}(t).
\ee
The norm of an eigenvector being positive, it remains to
impose the gap hypothesis in H3 to fit in the framework and we deduce
that the
$S$-matrix is unitarity, since $R=\un$. In this context,
the elements of the $S$-matrix describe the transitions between the
different levels between $t=-\infty$ and $t=+\infty$ in the adiabatic
 limit.

We now specify a little more our concern and consider a
three-level system, i.e. $H(t)=H^*(t)\in M_3(\C)$.
We assume that $H(t)$ satisfies the hypotheses of corollary \ref{PERCO}
with an extra parameter $\delta$ which we omit in the notation and
displays
two avoided crossings at $t_1<t_2$, as shown in figure \ref{adiab}.
\begin{figure}
\vspace{6cm}
\hspace{3cm}\special{picture fig7 scaled 500}
\caption{The pattern of avoided crossings in the adiabatic context.}
\label{adiab}
\end{figure}
The corresponding permutation $\sigma$ is given by
\be
  \sigma(1)=3,\;\;\; \sigma(2)=1, \;\;\; \sigma(3)=2.
\ee
By corollary \ref{PERCO}, we can compute asymptotically the elements
$s_{31},s_{12},s_{23}$ and
$s_{jj}$, $j=1,2,3$. Using the unitarity of the $S$-matrix, we can get
 some more information.
Introducing
\be
  \Gamma_j=\left|\mbox{Im }\int_{\zeta_j}e_j(z)dz\right| ,\;\;\; j=1,2,
\ee
where $\zeta_j$ is in the upper half plane, with the notation of
 section \ref{avoid}, it follows that
\be
  s_{31}=\ode \left(\e^{-(\Gamma_1 +\Gamma_2)/\eps}\right), \;\;\;
  s_{12}=\ode
  \left(\e^{-\Gamma_1/\eps}\right), \;\;\; s_{23}=\ode
  \left(\e^{-\Gamma_2/\eps}\right)
\ee
and
\be
  s_{jj}=1+\ode (\eps), \;\;\; j=1,2,3.
\ee
Expressing the fact that the first and second columns as well as the
second and third rows are
orthogonal, we deduce
\bea
  s_{21}&=&-\overline{s_{12}}\frac{s_{11}}{\overline{s_{22}}}
  \left(1+\ode \left(\e^{-2\Gamma_2/\eps}\right)\right)\\
  s_{32}&=&-\overline{s_{23}}\frac{s_{33}}{\overline{s_{22}}}
  \left(1+\ode \left(\e^{-2\Gamma_1/\eps}\right)\right).
\eea
Finally, the estimate in corollary \ref{PERCO} yields
\be
   s_{13}=\ode \left(\eps\e^{-|h|(e_2(\infty,\delta)-e_1(\infty,\delta))/
   \eps}\e^{-\Gamma_2/\eps}\right)
   =\ode \left(\e^{-(\Gamma_2+\Gamma_2+K)/\eps}\right)
\ee
where $K>0$, since we have that $\Gamma_j\ra 0$ as $\delta\ra 0$.
Hence we get
\be
  S=\pmatrix{s_{11}&s_{12}&\ode \left(\e^{-(\Gamma_2+\Gamma_2+K)/\eps}
  \right)\cr
              -\overline{s_{12}}\frac{s_{11}}{\overline{s_{22}}}
              \left(1+\ode \left(\e^{-2\Gamma_2/\eps}\right)\right)
              &s_{22}&s_{23}\cr
             s_{31}&-\overline{s_{23}}\frac{s_{33}}{\overline{s_{22}}}
             \left(1+\ode \left(\e^{-2\Gamma_1/\eps}\right)
             \right)&s_{33}}
\ee
where all $s_{jk}$ above can be computed asymptotically up to
exponentially small relative error, using
(\ref{imfor}).

The smallest asymptotically computable element $s_{13}$ describes the
transition from $e_1(-\infty,\delta)$ to
$e_3(+\infty,\delta)$. The result we get for this element is in agreement
with the rule
of the thumb claiming
that the transitions take place locally at the avoided crossings and can
be considered as
independent.
Accordingly, we can only estimate the smallest element of all, $s_{13}$,
 which
describes the transition from $e_3(-\infty,\delta)$ to
 $e_1(+\infty,\delta)$,
for which the avoided crossings are not encountered in "right order",
as discussed in \cite{hp}.
It is possible however to get an asymptotic expression for this element
in some cases.
When the unperturbed levels $e_2(z,0)$ and $e_3(z,0)$ possess a
degeneracy
point in $S_{\alpha}$ and when there exists a dissipative domain for the
index $3$ of the unperturbed
eigenvalues going above
this point, one can convince oneself that $s_{13}$ can be computed
asymptotically
for $\delta$ small enough, using the techniques presented
above.

Our second application is the study of the semi-classical scattering
properties of the multichannel
stationnary Schr\"odinger equation with energy above the potential
 barriers.
The relevant equation is then
\be\label{muls}
  -\eps^2\frac{d^2}{dt^2}\Phi(t)+V(t)\Phi(t)=E\Phi(t),
\ee
where $t\in\R$ has the meaning of a space variable, $\Phi(t)\in\C^m$
is
the wave function, $\eps\ra 0$ denotes
Planck's constant, $V(t)=V^*(t)\in M_m(\C)$ is the matrix of potentials
 and
the spectral parameter $E$ is kept fixed and large enough so that
\be\label{cosp}
  U(t)\equiv E-V(t)>0.
\ee
Introducing
\be
  \psi(t)=\pmatrix{\Phi(t)\cr i\eps \Phi(t)}\in \C^{2m},
\ee
we cast equation (\ref{muls}) into the equivalent form (\ref{schr})
for $\psi(t)$ with generator
\be\label{gess}
  H(t)=\pmatrix{\zer & \un\cr
                 U(t)& \zer}\in M_{2m}(\C).
\ee
It is readily verified that
\be
   H(t)=J^{-1}H^*(t)J
\ee
with
\be
  J=\pmatrix{\zer & \un\cr
              \un & \zer}.
\ee
Concerning the spectrum of $H(t)$, it should be remarked that if the
real and positive eigenvalues of $U(t)$, $k_j^2(t)$, $j=1,\cdots, m$
associated with
the eigenvectors $u_j(t)\in\C^m$  are
assumed to be distinct, i.e.
\be
  0<k_1^2(t)<k_2^2(t)<\cdots <k_m^2(t),
\ee
then the spectrum of the generator $H(t)$ given by (\ref{gess}) consists
in
$2m$ real distinct eigenvalues
\be\label{not1}
  -k_m(t)<-k_{m-1}(t)<\cdots <-k_1(t)<k_1(t)<k_2(t)<\cdots <k_m(t)
\ee
associated with the $2m$ eigenvectors
\bea\label{not2}
 & &\chi_j^{\pm}(t)=\pmatrix{u_j(t)\cr\pm k_j(t)u_j(t)}\in\C^{2m},
 \nonumber\\
 & &H(t)\chi_j^{\pm}(t)=\pm k_j(t)\chi_j^{\pm}(t).
\eea
We check that
\be\label{nvp}
  (\chi_j^{\pm}(0),\chi_j^{\pm}(0))_J=\pm2k_j(0)\|u_j(0)\|\neq 0,\;\;\;
   j=1,\dots ,m\,
\ee
where $\|u_j(0)\|$ is computed in $\C^m$, so that proposition \ref{PROS}
applies. Before dealing with its consequences,
we further explicit the structure of $S$.
Adopting the notation suggested by (\ref{not1}) and (\ref{not2}) we
write
\bea
  H(t)&=&\sum_{j=1}^nk_j(t)P_j^+(t)-\sum_{j=1}^nk_j(t)P_j^-(t)\\
  \label{decpm}
  \psi(t)&=&\sum_{j=1}^nc^+_j(t)\ffi_j^+(t)\e^{-i\int_0^tk_j(t')dt'/\eps}
  +
  \sum_{j=1}^nc^-_j(t)\ffi_j^-(t)\e^{i\int_0^tk_j(t')dt'/\eps}
\eea
and introduce
\be
  {\bf c}^{\pm}(t)=\pmatrix{c_1^{\pm}(t)\cr c_2^{\pm}(t)\cr \vdots \cr
  c_m^{\pm}(t)}\in \C^m.
\ee
Hence we have the block structure
\be\label{sm}
  S\pmatrix{{\bf c}^{+}(-\infty)\cr {\bf c}^{-}(-\infty)}\equiv
  \pmatrix{S_{++}&S_{+-}\cr S_{-+}&S_{--}}\pmatrix{{\bf c}^{+}(-\infty)
  \cr {\bf c}^{-}(-\infty)}=
  \pmatrix{{\bf c}^{+}(+\infty)\cr {\bf c}^{-}(+\infty)}
\ee
where $S_{\sigma \tau}\in M_{m}(\C)$, $\sigma, \tau\in\{+,-\}$.

Let us turn to the symmetry properties of $S$. We get from (\ref{nvp})
and
proposition \ref{PROS} that
\be
  \pmatrix{S_{++}&S_{+-}\cr S_{-+}&S_{--}}^{-1}=
  \pmatrix{\un & \zer \cr \zer & -\un}
  \pmatrix{S_{++}&S_{+-}\cr S_{-+}&S_{--}}^*
  \pmatrix{\un & \zer \cr \zer & -\un}=
  \pmatrix{S_{++}^*&-S_{-+}^*\cr -S_{+-}^*&
  S_{--}^*}.
\ee
In terms of the blocks $S_{\sigma \tau}$, this is equivalent to
\bea\label{222}
& &S_{++}S_{++}^*-S_{+-}S_{+-}^*=\un\\ \label{223}
& &S_{++}S_{-+}^*-S_{+-}S_{--}^*=\zer\\
& &S_{--}S_{--}^*-S_{-+}S_{-+}^*=\un .
\eea
The block $S_{++}$ describes the transmission coefficients associated
with a wave traveling
from the right and $S_{-+}$ describes the associated reflexion
coefficients.
Similarly, $S_{--}$ and $S_{+-}$ are related to the transmission and
reflexion
coefficients
associated with a wave incoming from the left. It should be noted that in
case of
equation (\ref{muls}) another convention is often used to define an
 $S$-matrix,
see \cite{f1}, for instance. This gives rise to a different $S$-matrix
 with
similar interpretation. However it is not difficult to establish a
one-to-one correspondence between the two definitions.
If the matrix of potentials $V(t)$ is real symmetric, we have further
 symmetry
in the $S$-matrix.
\begin{lem}\label{REPOT}
Let $S$ given by (\ref{sm}) be the $S$-matrix associated with
(\ref{muls}) under condition (\ref{cosp}). If we further assume
$V(t)=\overline{V(t)}$, then, taking $\ffi_j^{\pm}(0)\in \R^{2m}$,
$j=1,\cdots, m$, in
(\ref{decpm}), we get
\be
  S_{++}=\overline{S_{--}}\, ,\;\;\; S_{+-}=\overline{S_{-+}}.
\ee
\end{lem}
The corresponding results for the $S$-matrix defined in \cite{f1} are
derived
in \cite{mn}. The proof of this lemma can be found in appendix.
We consider now (\ref{muls}) the case
 $U(t)=U^*(t)=\overline{U(t)}\in M_2(\R)$, which
describes a two-channel Schr\"odinger equation. We assume that the
four-level generator
$H(t)$ displays three avoided crossings at $t_1<t_2$, two of which
take place at the same
point $t_1$, because of the symmetry of the eigenvalues, as in
figure \ref{semic}.
\begin{figure}
\vspace{7cm}
\hspace{3cm}\special{picture fig8 scaled 500}
\caption{The pattern of avoided crossings in the semiclassical context.}
\label{semic}
\end{figure}
By lemma \ref{REPOT}, it is enough to consider the blocks $S_{++}$ and
$S_{+-}$.
The transitions corresponding to elements of these blocks which we can
compute
asymptotically are from level $1^+$ to level $2^+$ and from level $2^-$
to level $1^+$.
They correspond to elements $s^{++}_{21}$ and $s^{+-}_{12}$ respectively.
 With the
notations
\be
  \Gamma_j=\left|\mbox{Im }\int_{\zeta_j}k_1(z)dz\right| ,\;\;\; j=1,2,
\ee
where $\zeta_j$ is in the upper half plane,
we have the estimates
\be\label{esss}
  s^{++}_{21}=\ode \left(\e^{-\Gamma_1/\eps}\right), \;\;\;
  s^{+-}_{12}=\ode \left(\e^{-(\Gamma_1 +\Gamma_2)/\eps}\right),\;\;\;
  s^{++}_{jj}=1+\ode (\eps), \;\;\; j=1,2.
\ee
It follows from (\ref{223}) and lemma \ref{REPOT} that the matrix
$S_{++}S_{+-}^T$
is symmetric. Hence
\be\label{resly}
  s^{++}_{11}s^{+-}_{21}+s^{++}_{12}s^{+-}_{22}=
  s^{++}_{21}s^{+-}_{11}+s^{++}_{22}s^{+-}_{12},
\ee
whereas we get from (\ref{222})
\be\label{from}
  s^{++}_{11}\overline{s^{++}_{21}}+s^{++}_{12}\overline{s^{++}_{22}}=
  s^{+-}_{11}\overline{s^{+-}_{21}}+s^{+-}_{12}\overline{s^{+-}_{22}}.
\ee
The only useful estimate we get with corollary \ref{PERCO} is
\be
  s^{+-}_{22}=\ode \left(\e^{-(\Gamma_1 +\Gamma_2+K)/\eps}\right),\;\;\;
   K>0,
\ee
which yields together with (\ref{esss}) in (\ref{resly})
\be
  s^{+-}_{21}=s^{++}_{21}s^{+-}_{11}/s^{++}_{11}+\ode
  \left(\e^{-(\Gamma_1 +\Gamma_2)/\eps}\right).
\ee
Thus, from (\ref{from}) and (\ref{sjksup}) for $s^{+-}_{11}$,
\be
  s^{++}_{12}=-\overline{s^{++}_{21}}\frac{s^{++}_{11}}
  {\overline{s^{++}_{22}}}
  \left(1+\ode \left(\e^{-\kappa/\eps}\right)\right),
\ee
with
\be\label{bbeh}
  0<\kappa<\min(\Gamma_1,\Gamma_2).
\ee
Summarizing, we have
\be
  S_{++}=\pmatrix{s^{++}_{11}&-\overline{s^{++}_{21}}\frac{s^{++}_{11}}
  {\overline{s^{++}_{22}}}
  \left(1+\ode \left(\e^{-\kappa/\eps}\right)\right)\cr
           s^{++}_{21}&s^{++}_{22}}
\ee
and
\be
  S_{+-}=\pmatrix{\ode \left(\e^{-\kappa/\eps}\right) &s^{+-}_{12}\cr
           \ode \left(\e^{-\kappa/\eps}\right) &\ode \left(
           \e^{-(\Gamma_1 +\Gamma_2+K)/\eps}\right)},
\ee
where all elements $s^{\sigma\tau}_{jk}$ can be asymptotically computed
up to exponentially
small relative corrections using (\ref{imfor}). We get no information on
the first
column of $ S_{+-}$ but the estimate (\ref{sjksup}) where necessarily,
(\ref{bbeh}) holds.
However, if there exists one or several
other dissipative domains for certain indices, it is then possible to get
 asymptotic formulas for
the estimated terms.

\appendix
\section{Proof of proposition
\protect{\ref{PROS}}}
\renewcommand{\theequation}{\Alph{section}.\arabic{equation}}
\setcounter{equation}{0}

A direct consequence of the property
\be\label{sead}
  H(t)=H^{\#}(t)=J^{-1}H^*(t)J
\ee
is the relation $\sigma(H(t))=\overline{\sigma(H(t))}$. Thus, if
$\sigma(H(0))\subset \R$, then $\sigma(H(t))\subset \R$, for all
$t\in\R$,
since the analytic eigenvalues are assumed to be distinct and
nondegenerate
for all $t\in\R$.
Let $e_j(0)$ be the eigenvalue associated with $\ffi_j(0)$. Then, due
to the property $H(0)=H^{\#}(0)$
\be
  (\ffi_j(0),H(0)\ffi_k(0))_J=e_k(0)(\ffi_j(0),\ffi_k(0))_J=
  \overline{e_j(0)}(\ffi_j(0),\ffi_k(0))_J,
\ee
for any $j,k=1,\cdots,n$. For $j=k$ we get from the assumption
$(\ffi_j(0),\ffi_j(0))_J\neq 0$ that $e_j(0)\in\R$ and from the fact
that
the eigenvalues of $H(0)$ are distinct
\be
  (\ffi_j(0),\ffi_k(0))_J=0,\;\;\;j\neq k.
\ee
The resulting reality of $e_j(t)$ for all $t\in\R$ and $j=1,\cdots,n$
 yields
together with (\ref{sead})
\be
  P_j(t)=J^{-1}P_j^*(t)J.
\ee
Hence, using the fact that the $P_j^*$ are projectors,
\bea
  K(t)&=&\sum_{j=1}^n{P_j}'(t)P_j(t)=
  \sum_{j=1}^n(J^{-1}P_j^*(t)J)'J^{-1}P_j^*(t)J=J^{-1}\sum_{j=1}^n
  {P_j^*}'(t)P_j^*(t)J\nonumber\\
  &=&-J^{-1}\sum_{j=1}^nP_j^*(t){P_j^*}'(t)J=-J^{-1}K^*(t)J.
\eea
Let $\Phi,\Psi\in\C^n$ and $W(t)$ be defined by (see (\ref{par}))
\be
  W'(t)=K(t)W(t), \;\;\; W(0)=\un.
\ee
Then we have
\bea
  (W(t)\Phi,W(t)\Psi)_J'&=&\bra W'(t)\Phi |JW(t)\Psi\ket+
  \bra W(t)\Phi |JW'(t)\Psi\ket\nonumber \\
  &=&\bra K(t)W(t)\Phi |JW(t)\Psi\ket+\bra W(t)\Phi |JK(t)W(t)\Psi\ket
  \nonumber \\
  &=&\bra W(t)\Phi |J(J^{-1}K^*(t)J+K(t))W(t)\Psi\ket\equiv 0.
\eea
Thus, in the indefinite metric, the scalar products of the eigenvectors
of $H(t)$,
$\ffi_j(t)=W(t)\ffi_j(0)$ (see (\ref{defeig})), are constants
\be
  (\ffi_j(t),\ffi_k(t))_J\equiv(\ffi_j(0),\ffi_k(0))_J.
\ee
We can then normalize the $\ffi_j(0)$ in such a way that
\be
  (\ffi_j(t),\ffi_k(t))_J=(\ffi_j(0),\ffi_k(0))_J=\delta_{jk}\rho_j
\ee
with $\rho_j\in\{+1,-1\}$.
Let $\psi(t)$ and $\chi(t)$ be two solutions of (\ref{schr}). By a
argument similar
to the one above using (\ref{sead}), we deduce
\be
  (\chi(t),\psi(t))_J\equiv(\chi(0),\psi(0))_J.
\ee
Inserting the decompositions
\bea
  \psi(t)&=&\sum_{j=1}^nc_j(t)\e^{-i\int_0^te_j(t')dt'/\eps }\ffi_j(t)\\
  \chi(t)&=&\sum_{j=1}^nd_j(t)\e^{-i\int_0^te_j(t')dt'/\eps}\ffi_j(t)
\eea
in this last identity yields
\bea\label{refi}
  & &\sum_{j,k=1}^n\overline{d}_k(t){c}_j(t)(\ffi_k(t),\ffi_j(t))_J
  \e^{i\int_0^t(e_k(t')-e_j(t'))/\eps dt'}=
  \sum_{j}^n\overline{d}_j(t)\rho_j{c}_j(t)\nonumber\\
  & &\equiv\sum_{j=1}^n\overline{d}_j(0)\rho_j{c}_j(0)
  =\sum_{j=1}^n\overline{d}_j(\pm\infty)\rho_j{c}_j(\pm\infty).
\eea
Since the initial conditions for the coefficients
\be
  {c}_j(-\infty)=\delta_{jk}\, ,\;\;\;{d}_j(-\infty)=\delta_{jl}
\ee
imply
\be
  {c}_j(+\infty)=s_{jk}\, ,\;\;\;{d}_j(+\infty)=s_{jl},
\ee
we get from (\ref{refi}), introducing the matrix
$R=\mbox{diag }(\rho_1,\rho_2,\cdots, \rho_n)\in M_n(\C)$,
\be
 R=S^*RS ,
\ee
which is equivalent to the assertion $S^{-1}=RS^*R$.\ep

\section{Proof of lemma
\protect{\ref{REPOT}}}
\setcounter{equation}{0}

Let $G=G^*=G^{-1}$ be given in block structure by
\be
  G=\pmatrix{\un & \zer \cr \zer & -\un}\in M_{2m}(\C)
\ee
and $H(t)$ be given by (\ref{gess}) with $U(t)=\overline{U(t)}=U^*(t)$.
Since
\be\label{sym}
  GH(t)G=-H(t)\, , \;\;\; \overline{H(t)}=H(t)\, ,
\ee
and the eigenvalues of $H(t)$ are real, it is readily verified that
\be
 GP_j^{\pm}(t)G=P_j^{\mp}(t)\, , \;\;\; \overline{P_j^{\pm}(t)}=
 P_j^{\pm}(t)\, , \; j=1,\cdots, m.
\ee
Hence
\be
  K(t)=\sum_{j=1 \atop \tau=\pm}^m{P_j^{\tau}}'(t)P_j^{\tau}(t)=
  \overline{K(t)}=GK(t)G,
\ee
from which follows that the solution $W(t)$ of
\be
  W'(t)=K(t)W(t)\, , \;\;\; W(0)=\un
\ee
satisfies
\be
 W(t)=\overline{W(t)}=GW(t)G.
\ee
As the matrix of potentials $U(0)$ is real symmetric, its eigenvectors
$u_j(0)$ may be chosen real, so that we can assume that
\be
  \ffi_j^{\pm}(0)=\pmatrix{u_j(0)\cr \pm k_j(0)u_j(0)}\in\R^{2m}.
\ee
Thus it follows from the foregoing that
\be\label{real}
  \ffi_j^{\pm}(t)=W(t)\ffi_j^{\pm}(0)\in\R^{2m}\;\;\; \forall t\in\R
\ee
and satisfies
\be\label{gsy}
  G\ffi_j^{\pm}(t)=GW(t)GG\ffi_j^{\pm}(0)=W(t)G\ffi_j^{\pm}(0)
  =\ffi_j^{\mp}(t).
\ee
Finally, the main consequence of (\ref{sym}) is that if $\psi(t)$ is a
solution
of
\be
  i\eps \psi'(t)=H(t)\psi(t),
\ee
then $\ffi(t)=G\overline{\psi(t)}$ is another solution, as easily
verified.
Thus we can write with (\ref{decpm}),(\ref{real}) and (\ref{gsy}) that
\bea
  \ffi(t)&=&\sum_{j=1}^md^+_j(t)\ffi_j^+(t)\e^{-i\int_0^tk_j(t')dt'/\eps}
  +
  \sum_{j=1}^md^-_j(t)\ffi_j^-(t)\e^{i\int_0^tk_j(t')dt'/\eps}\nonumber\\
  &=&\sum_{j=1}^m\overline{c^+_j(t)}\ffi_j^-(t)\e^{i\int_0^tk_j(t')dt'/
  \eps}+
  \sum_{j=1}^m\overline{c^-_j(t)}\ffi_j^+(t)\e^{-i\int_0^tk_j(t')dt'/
  \eps},
\eea
i.e.
\bea
  d^+_j(t)&=&\overline{c^-_j(t)}\nonumber\\
  d^-_j(t)&=&\overline{c^+_j(t)}\, ,\;\;\; \forall j=1,\cdots, m\, ,
  \;\forall
  t\in\R.
\eea
Finally, using the definition (\ref{sm}) and the above property for
$t=\pm\infty$, we get for any ${\bf d}^{\pm}(-\infty)\in\C^m$
\be
  \pmatrix{{\bf d}^+(+\infty)\cr {\bf d}^-(+\infty)}=
  \pmatrix{S_{++}&S_{+-}\cr S_{-+}&S_{--}}\pmatrix{{\bf d}^{+}(-\infty)
  \cr {\bf d}^{-}(-\infty)}=\pmatrix{\overline{S_{--}}&\overline{S_{-+}}
  \cr
  \overline{S_{+-}}&\overline{S_{++}}}\pmatrix{{\bf d}^{+}(-\infty)
  \cr {\bf d}^{-}(-\infty)},
\ee
from which the result follows. \ep

\end{document}